\documentclass[prx,twocolumn,amsmath,amssymb,superscriptaddress]{revtex4-1}

\usepackage{amsmath}    
\usepackage{graphicx}   
\usepackage{verbatim}   
\usepackage{color}      
\usepackage{subfigure}  
\usepackage{hyperref}   
\usepackage{multirow}
\usepackage{float}
\usepackage{braket}
\usepackage{siunitx}
\usepackage{bbm}

\allowdisplaybreaks

\newcommand{\nn}{\nonumber}

\begin{document}

\title{Superconducting Quantum Simulator for Topological Order and the Toric Code}
\author{Mahdi Sameti}
\affiliation{Institute of Photonics and Quantum Sciences, Heriot-Watt University Edinburgh EH14 4AS, United Kingdom}
\author{Anton Poto\v{c}nik}
\affiliation{Department of Physics, ETH Z\"urich, CH-8093 Z\"urich, Switzerland}
\author{Dan E. Browne}
\affiliation{Department of Physics and Astronomy, University College London, Gower Street, London WC1E 6BT, United Kingdom}
\author{Andreas Wallraff}
\affiliation{Department of Physics, ETH Z\"urich, CH-8093 Z\"urich, Switzerland}
\author{Michael J. Hartmann}
\affiliation{Institute of Photonics and Quantum Sciences, Heriot-Watt University Edinburgh EH14 4AS, United Kingdom}
\date{\today}

\begin{abstract}
Topological order is now being established as a central criterion for characterizing and classifying ground states of condensed matter systems and complements categorizations based on symmetries. Fractional quantum Hall systems and quantum spin liquids are receiving substantial interest because of their intriguing quantum correlations, their exotic excitations and prospects for protecting stored quantum information against errors. Here we show that the Hamiltonian of the central model of this class of systems, the Toric Code, can be directly implemented as an analog quantum simulator in lattices of superconducting circuits.
The four-body interactions, which lie at its heart, are in our concept realized via Superconducting Quantum Interference Devices (SQUIDs) that are driven by a suitably oscillating flux bias. All physical qubits and coupling SQUIDs can be individually controlled with high precision. Topologically ordered states can be prepared via an adiabatic ramp of the stabilizer interactions. Strings of qubit operators, including the stabilizers and correlations along non-contractible loops, can be read out via a capacitive coupling to read-out resonators. Moreover, the available single qubit operations allow to create and propagate elementary excitations of the Toric Code and to verify their fractional statistics. The architecture we propose allows to implement a large variety of many-body interactions and thus provides a versatile analog quantum simulator for topological order and lattice gauge theories.
\end{abstract}

\maketitle


Topological phases of quantum matter \cite{Stormer99,Balents10} are of scientific interest for their intriguing quantum correlation properties, exotic excitations with fractional and non-Abelian quantum statistics, and their prospects for physically protecting quantum information against errors \cite{Nayak08}.

The model which takes the central role in the discussion of topological order is the Toric Code \cite{Kitaev2003,Kitaev09,Fowler12}, which is an example of a $\mathbb{Z}_2$ lattice gauge theory. It consists of a two-dimensional spin lattice with quasi local four-body interactions between the spins and features $4^{g}$ degenerate, topologically ordered ground states for a lattice on a surface of genus $g$. The topological order of these $4^{g}$ ground states shows up in their correlation properties. The topologically ordered states are locally indistinguishable (the reduced density matrices of a single spin are identical for all of them) and only show differences for global properties (correlations along non-contractible loops). Moreover local perturbations cannot transform these states into one another and can therefore only excite higher energy states that are separated by a finite energy gap. 

In equilibrium, the Toric Code ground states are therefore protected against local perturbations that are small compared to the gap, which renders them ideal candidates for storing quantum information \cite{Kitaev2003}. Yet for a self-correcting quantum memory, the mobility of excitations needs to be suppressed as well, which so far has only been shown to be achievable in four or more lattice dimensions \cite{Dennis02,Bombin13}.
Excitations above the ground states of the Toric Code are generated by any string of spin rotations with open boundaries. The study of these excitations, called anyons, is of great interest as the neither show bosonic nor fermionic but fractional statistics \cite{Kitaev2003,Kitaev09}.

\begin{figure}[t]
\includegraphics[width=0.95\columnwidth]{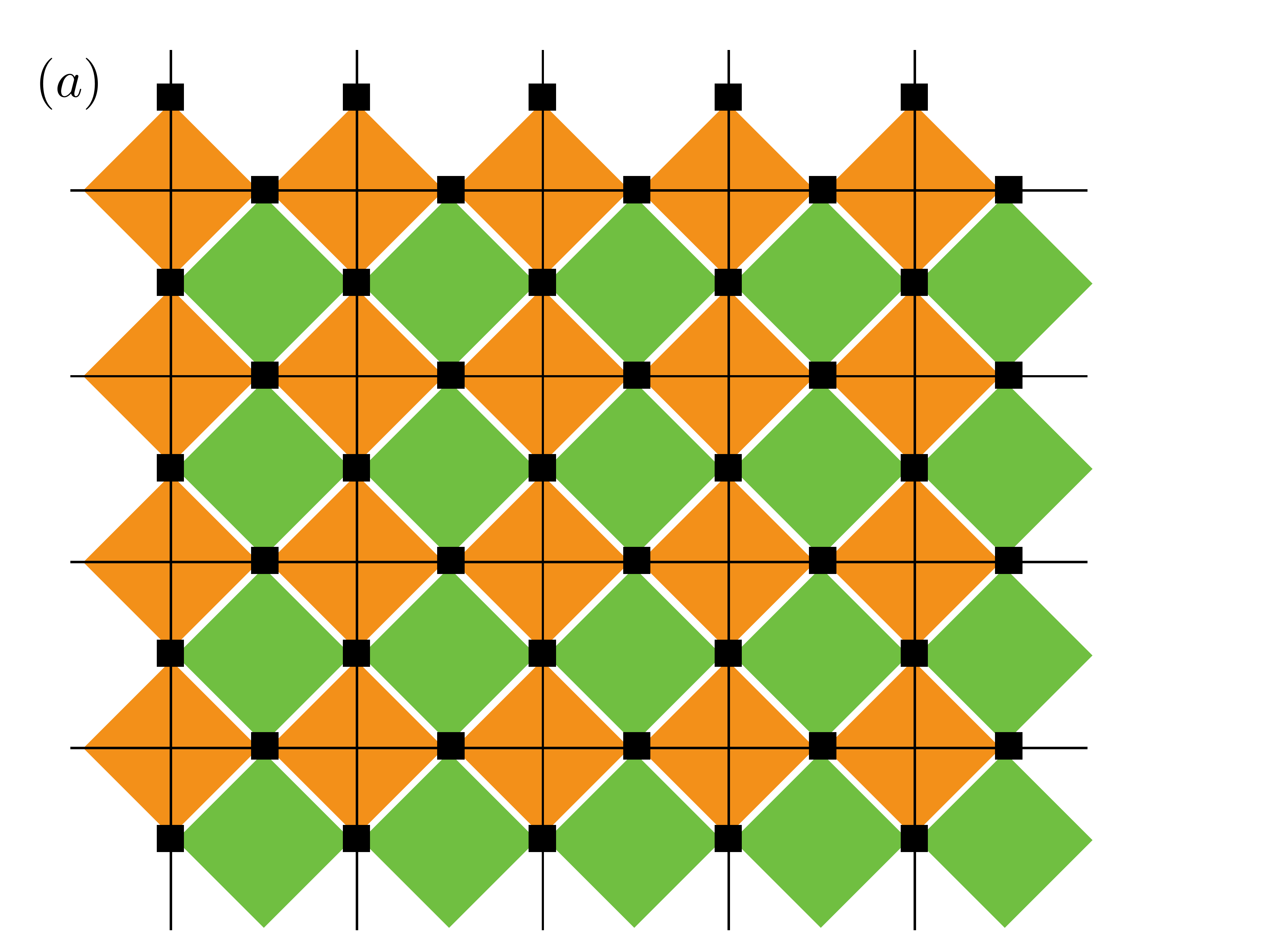}\\
\vspace{5mm}
\includegraphics[width=0.96\columnwidth]{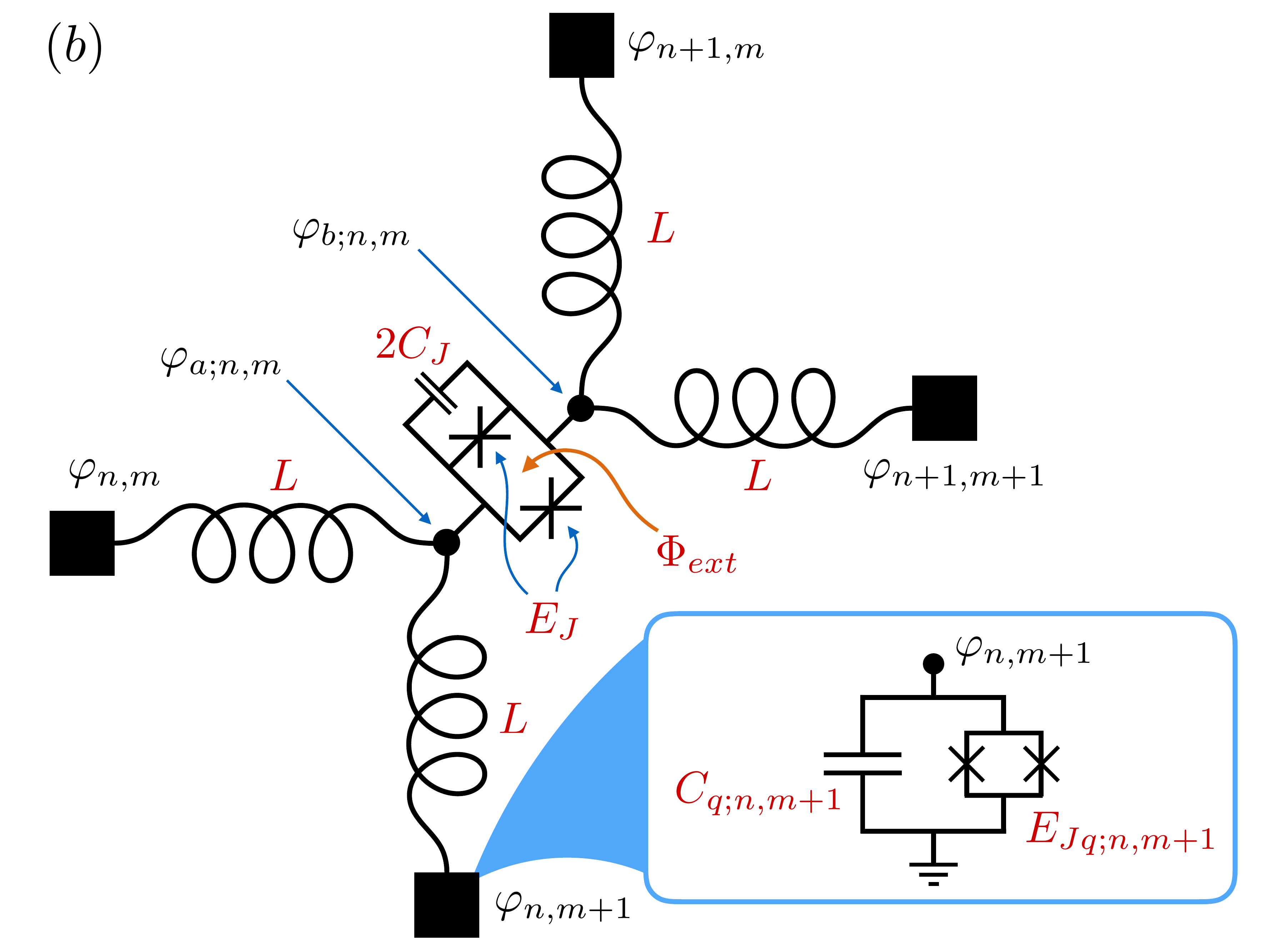}
\caption{\label{fig:lattice-main} Toric Code lattice and its implementation. {\bf (a)} Lattice for the Toric Code. Small rectangles on the lattice edges indicate spin-1/2 systems (qubits). Star interactions $A_{s}$ are marked with orange diamonds and plaquette interactions $B_{p}$ with green diamonds.
The transition frequencies of the qubits form a periodic pattern in both lattice directions such that no nearest neighbors and no next nearest neighbors of spins have the same transition frequency and therefore all four spins at the corners of a physical cell have different transition frequencies.
{\bf (b)} Circuit of a physical cell. The rectangles at the corners represent qubits with different transition frequencies, see inset for the circuit of each qubit. Qubits $(n,m)$ and $(n,m+1)$ [$(n+1,m)$ and $(n+1,m+1)$] are connected via the inductances $L$ to the node $(a;n,m)$ [$(b;n,m)$]. The nodes $(a;n,m)$ and $(b;n,m)$ in turn are connected via a dc-SQUID with an external magnetic flux $\Phi_{ext}$ threaded through its loop. The Josephson junctions in the dc-SQUID have Josephson energies $E_{J}$ and capacitances $C_{J}$, which include shunt capacitances.}
\end{figure}

In two dimensions, the Toric Code can be visualized as a square lattice where the spin-1/2 degrees of freedom are placed on the edges, see Fig.~\ref{fig:lattice-main}a. 
Stabilizer operators can be defined for this lattice as products of $\sigma^x$-operators for all four spins around a star (orange diamond in Fig.~\ref{fig:lattice-main}a), i.e. $A_s = \prod_{j\in \text{star}(s)} \sigma_j^x$, and products of $\sigma^y$-operators for all four spins around a plaquette (green diamond in Fig.~\ref{fig:lattice-main}a), i.e. $B_p = \prod_{j \in \text{plaq}(p)} \sigma_j^y$. The Hamiltonian for this system reads,
\begin{equation} \label{eq:toric-code-ham}
  H_{\text{TC}} = -J_s \sum_s A_s -J_p \sum_p B_p,
\end{equation}
where $J_{s} (J_{p})$ is the strength of the star (plaquette) interactions and the sums $\sum_{s}$ ($\sum_{p}$) run over all stars (plaquettes) in the lattice \cite{convention}.

Despite the intriguing perspectives for quantum computation and for exploring strongly correlated condensed matter physics that a realization of the Toric Code Hamiltonian would offer, progress towards its implementation has to date been inhibited by the difficulties in realising clean and strong four-body interactions while at the same time suppressing two-body interactions to a sufficient extent. Realisations of topologically encoded qubits have been reported in superconducting qubits \cite{Ioffe02,Gladchenko09}, photons \cite{Yao12} and trapped ions \cite{Nigg14}. The Toric Code is also related to a model of geometrically frustrated spins on a honeycomb lattice \cite{Kitaev06}. Yet, in contrast to the Toric Code itself, this Honeycomb model only features two-body interactions, which can naturally occur in spin lattices, and evidence for its realization in iridates \cite{Chun15} and ruthenium-based materials \cite{Banerjee16} has been found in neutron scattering experiments.

There is moreover an important difference in the use and purpose of an implementation of the Hamiltonian (\ref{eq:toric-code-ham}) compared to error detection via stabilizer measurements as pursued in \cite{Yao12,Kelly14,Riste14,Corcoles15,Blumoff16}. In particular, two-body interactions are all that is needed for arbitrary stabiliser measurements, while lattice gauge theories \cite{Marcos13,Marcos14,Hafezi13,Brennen15} and Hamiltonians such as (\ref{eq:toric-code-ham}) require $k$-body (with $k>2$) effective interactions. We note that the realisation of $k$-body interactions is also important in quantum annealing architectures, where implementations have been proposed utilising local interactions and many-body constraints \cite{Lechner15,Chancellor16,Leib16}.  We anticipate that the approach we describe here may also be relevant for quantum annealing, however our focus in this paper is on the realisation of Hamiltonians supporting topological order.

Superconducting circuits have recently made tremendous advances in realizing engineered quantum dynamics for quantum simulation \cite{Chen14,Barends15,Salathe15,Eichler15,OMalley16} and quantum information processing \cite{Reed12,Steffen13,Barends14,Kelly14,Riste14,Corcoles15,Barends16,Ofek16}. 
Their quantum features originate from integrated Josephson junctions that behave like ideal nonlinear inductors. They are typically built as SQUIDs that can be tuned and driven via magnetic fluxes. Currently superconducting circuits are assembled in ever larger connected networks \cite{Fitzpatrick16,Steffen13,Barends14,Kelly14,Riste14,Barends16}.
Here we make use of these exquisite properties of superconducting circuits for a scheme implementing the Hamiltonian of the two-dimensional Toric Code in a direct way by creating four-body interactions of the form $A_s = \prod_{j\in \text{star}(s)} \sigma_j^x$ and $B_p = \prod_{j \in \text{plaq}(p)} \sigma_j^y$ between the respective qubits and fully suppressing all two-body interactions. 

Superconducting circuits are an ideal platform for experimentally exploring topological order since high precision control and measurements of both, individual qubits as well as multi-qubit clusters have been demonstrated \cite{Filipp09,DiCarlo10,Kelly14,Salathe15,Roushan16} . Importantly, this platform is also suitable for realizing periodic boundary conditions, since conductors can cross each other via air-bridges \cite{Abuwasib13,Chen14a}. This property, which is essential for exploring topological order, is so far not accessible in other platforms unless one resorts to digitized approaches. 

Our approach is designed to keep the necessary circuitry as simple as possible and only employs standard superconducting qubits that are coupled via dc-SQUIDs  \cite{Jin13,Neumeier13}.
To generate the desired interactions we make use of the nonlinear nature of the coupling SQUIDs and the possibilities to bias them with constant and oscillating magnetic fluxes. In particular the four-body stabilizer interactions of the Toric Code are generated by driving the SQUIDs with suitably oscillating magnetic fluxes. They can therefore be switched on and off at desired times or even be smoothly tuned in strength.
In this way the Toric Code Hamiltonian is realized in a frame where the qubits rotate at their transition frequencies. Importantly, the transformation between the rotating and the laboratory frame is composed of single qubit unitaries only, so that the entanglement properties of the states are identical in both frames. Yet for implementing the Toric Code Hamiltonian (\ref{eq:toric-code-ham}), considering a rotating frame has the pivotal advantages that the effective transition frequencies of the individual spins become arbitrarily small and that the approach becomes largely insensitive to disorder in the lattice. In contrast to earlier approaches to multi-body interactions based on gate sequences \cite{Weimer10,Becker13,Tanamoto13,Mezzacapo14}, the four-body interactions in our approach are directly implemented as an analog quantum simulator and are therefore time-independent, which avoids any errors routed in the Trotter sequencing. Moreover, our approach suppresses two-body interactions to an extent that they are negligible compared to the four-body interactions.

A minimal two-dimensional lattice covering the surface of a torus requires eight qubits connected via superconducting wires such that the lattice has periodic boundary conditions in both directions and is thus realizable with available technology \cite{Steffen13,Kelly14,Riste14}.
We show that the topologically ordered states of the Toric Code can be prepared via an adiabatic sweep from a product state for experimentally realizable parameters and that measurements of individual qubits and qubit-qubit correlations can be performed to demonstrate the topological order of these states. Moreover the robustness of the topologically ordered states against local perturbation that would change the energy of the system by a limited amount can be verified experimentally.

\section{The Physical Lattice}
\label{sec:physical-lattice}
For realizing the Toric Code lattice, star and plaquette interactions, indicated by orange and green diamonds in Fig.~\ref{fig:lattice-main}a, are both realized via the circuit shown in Fig.~\ref{fig:lattice-main}b but with different external fluxes $\Phi_{ext}$ applied to the coupling SQUID. The lattice resulting from the orange and green diamonds in Fig.~\ref{fig:lattice-main}a with spins on the lattice vertices rather than edges is a convenient representation introduced in \cite{Bombin07}, see also \cite{Tomita14,Gambetta15}. The small black squares in Figs. \ref{fig:lattice-main}a and \ref{fig:lattice-main}b indicate spins formed by the two lowest energy eigenstates of superconducting qubits.
For the transition frequencies between these two lowest states, we assume a periodic 
pattern in both lattice directions such that no nearest neighbors and no next nearest neighbors of qubits have the same transition frequency and hence all qubits in each four-spin cell have different transition frequencies, see Sec.~\ref{sec:parameters} for details.

For each cell the four qubits on its corners are connected by a dc-SQUID formed by two identical Josephson junctions with Josephson energy $E_{J}$, see Fig \ref{fig:lattice-main}b (an asymmetry between the junctions causes a negligible perturbation only).
The dc-SQUID is threaded by a time dependent external flux, $\Phi_{ext}$ 
and its ports are connected to the four qubits via four inductors, each with inductance $L$. There are thus auxiliary nodes at both ports of the SQUID. As we show in Sec.~\ref{sec:3}, these auxiliary nodes $(a;n,m)$ and $(b;n,m)$ can be eliminated from the description while, together with the flux bias $\Phi_{ext}$ on the SQUID, they mediate the desired four-body interactions of the stabilizers. 

\subsection{Hamiltonian of the physical lattice}

We describe the lattice in terms of the phases $\varphi_{j}$ and their conjugate momenta $\pi_{j}$ for each node, which obey the commutation relations $\left[\varphi_{j},\pi_{l}\right] = i \hbar\delta_{j,l}$. The nodes are labelled by composite indices $j = (n,m)$, where $n$ labels the diagonals extending from top left to bottom right and $m$ labels the qubits along each diagonal in Fig.~\ref{fig:lattice-main}a.  

Each qubit is characterized by a Josephson energy $E_{Jq;j}$ and a charging energy $E_{Cq;j} = e^{2}/(2 C_{q;j})$, where $e$ is the elementary charge and $C_{q;j}$ the qubit's capacitance. Including a contribution $4 E_L \varphi_{j}^2$ that arises from the coupling to its four adjacent auxiliary nodes
via the inductances $L$, the Hamiltonian of the qubit at node $j$ reads, 
\begin{equation} \label{eq:qubitham}
H_{q;j} =  4 E_{Cq;j} \hbar^{-2} \pi_{j}^{2} - E_{Jq;j} \cos(\varphi_{j}) + 4 E_L \varphi_{j}^2,
\end{equation}
where $E_L = \phi_0^2/(2 L)$ and $\phi_0 = \hbar/(2 e)$ is the rescaled quantum of flux.
We consider transmon qubits \cite{Koch07,Steffen13,Barends14,Kelly14,Riste14,Barends15,Salathe15} because of their favorable coherence properties and moderate charging energies, c.f. appendix \ref{sec:first-order-corrs-adiabat}. 

For the description of the dc-SQUIDs it is useful to consider the modes $\varphi_{\pm;j} = \varphi_{a;j} \pm \varphi_{b;j}$, where $\varphi_{a/b;j}$ is the phase at the auxiliary node $(a/b;n,m)$. The modes $\varphi_{+;j}$ behave as harmonic oscillators described by the Hamiltonians
$H_{+;j} = 4 E_{C+} \hbar^{-2} \pi_{+;j}^{2} + E_L \varphi_{+;j}^2$, 
where $\pi_{\pm;j}$ are the canonically conjugate momenta to $\varphi_{\pm;j}$ and $E_{C+} = e^{2}/C_{g}$, with $C_g$ the capacitance between
each of the nodes $(a;j)$ or $(b;j)$ and ground.
The modes $\varphi_{-;j}$ in turn are influenced by the dc-SQUID. Their Hamiltonians read
\begin{equation}
H_{-;j} = 4 \frac{E_{C-}}{\hbar^{2}} \pi_{-;j}^{2} + E_L \varphi_{-;j}^2 - 2 E_J \cos\left(\frac{\varphi_{ext}}{2}\right) \cos(\varphi_{-;j}),
\end{equation}
where $\varphi_{ext} = \Phi_{ext}/\phi_{0}$ is the phase associated to the external flux through the SQUID loop, $E_{J}$ the Josephson energy of each junction in the SQUID and $E_{C-} = e^{2}/ 2 C_{J}$ with $C_{J}$ the capacitance of one junction in the SQUID, including a shunt capacitance parallel to the junction.

We choose the indices $j=(n,m)$ for the auxiliary nodes such that node $(a;n,m)$ couples to the qubits at nodes $(n,m)$ and $(n,m+1)$ whereas node $(b;n,m)$ couples to the qubits at nodes $(n+1,m)$ and $(n+1,m+1)$, see Fig.~\ref{fig:lattice-main}b.
The interaction terms between the qubit variables $\varphi_{j}$ and the SQUID modes $\varphi_{\pm;j}$ due to the inductances $L$ thus read,
\begin{equation} \label{eq:qubit-squid-coupling}
H_{L;j} = - E_L \varphi_{q+;j} \varphi_{+;j} - E_{L} \varphi_{q-;j} \varphi_{-;j}
\end{equation}
where $\varphi_{q\pm;j} = \varphi_{n,m}+\varphi_{n,m+1} \pm \varphi_{n+1,m} \pm \varphi_{n+1,m+1}$.
Here, inductive couplings via mutual inductances are an alternative option \cite{Kafri16}.
The full Hamiltonian of the lattice is therefore given by
\begin{equation}\label{eq:ham-full}
\mathcal{H} = \sum_{j=(1,1)}^{(N,M)} \left[H_{q;j} + H_{+;j} + H_{-,j} + H_{L;j} \right]
\end{equation}
A detailed derivation of the Hamiltonian $\mathcal{H}$, starting from a Lagrangian for the circuit, is provided in appendix \ref{sec:lagrangian}.
We now turn to explain how the four-body interactions of the stabilizers emerge in our architecture.

\section{Stabilizer operators} \label{sec:3}
The guiding idea for our approach comes from the following approximate consideration (We present a quantitatively precise derivation in Sec.~\ref{sec:elim-squid}).
In each single cell, as sketched in Fig.~\ref{fig:lattice-main}b, the auxiliary nodes $(a;n,m)$ and $(b;n,m)$ have very small capacitances with respect to ground and therefore cannot accumulate charge. As a consequence the node phases $\varphi_{a;n,m}$ and $\varphi_{b;n,m}$ are not dynamical degrees of freedom but instantly follow the phases at nodes $(n,m)$ and $(n,m+1)$ or $(n+1,m)$ and $(n+1,m+1)$ as given by Kirchoff's current law saying that the sum of all currents into and out of such a node must vanish. Therefore the Josephson energy of the coupling SQUID, $2 E_J \cos(\varphi_{ext}/2) \cos(\varphi_{a;n,m}-\varphi_{b;n,m})$, becomes a nonlinear function of the node phases $\varphi_{n,m}$, $\varphi_{n,m+1}$, $\varphi_{n+1,m}$ and $\varphi_{n+1,m+1}$, which contains four-body terms.

The external flux applied through the SQUID is then chosen such that it triggers the desired four-body interactions forming the stabilizers. To this end it is composed of a constant part with associated phase $\varphi_{dc}$ and an oscillating part associated with phase amplitude $\varphi_{ac}$,
\begin{equation} \label{eq:drive}
 \varphi_{ext}(t) = \varphi_{dc} + \varphi_{ac} F(t),
\end{equation}
where the oscillation frequencies of $F(t)$ are linear combinations of the qubit transition frequencies such that the desired four-body spin interactions of the Hamiltonian (\ref{eq:toric-code-ham}) are enabled and all other interaction terms are suppressed. Moreover we choose 
$\varphi_{dc} = \pi$ to suppress the critical current $I_{c} = 2 E_{J} \phi_{0}^{-1} \cos ( \varphi_{dc}/2)$ of the dc-SQUID \cite{Nakamura99} and thus unwanted interactions of excitations between qubits at opposite sides of it, c.f. Fig.~\ref{fig:lattice-main}b. Assuming furthermore that the oscillating part of the flux bias is small compared to the constant part, we approximate $\cos(\varphi_{ext}/2) \approx - \varphi_{ac} F(t)/2$. 

The above consideration uses approximations and only serves us as a guide.
To obtain a quantitatively precise picture, we here take the small but finite capacitances of the Josephson junctions in the SQUID into account and derive the stabilizer interactions in a fully quantum mechanical approach. Our derivation makes use of methods for generating effective four-body Hamiltonians in the low energy sector of a two-body Hamiltonian \cite{Kitaev06,Bravyi08,Koenig10,Brell10} by employing a Schrieffer-Wolff transformation \cite{Schrieffer66} and combines these with multi-tone driving in order to single out specific many-body interactions while fully suppressing all two-body interactions.

\begin{figure}[h!]
\centering
\includegraphics[width=.9\columnwidth]{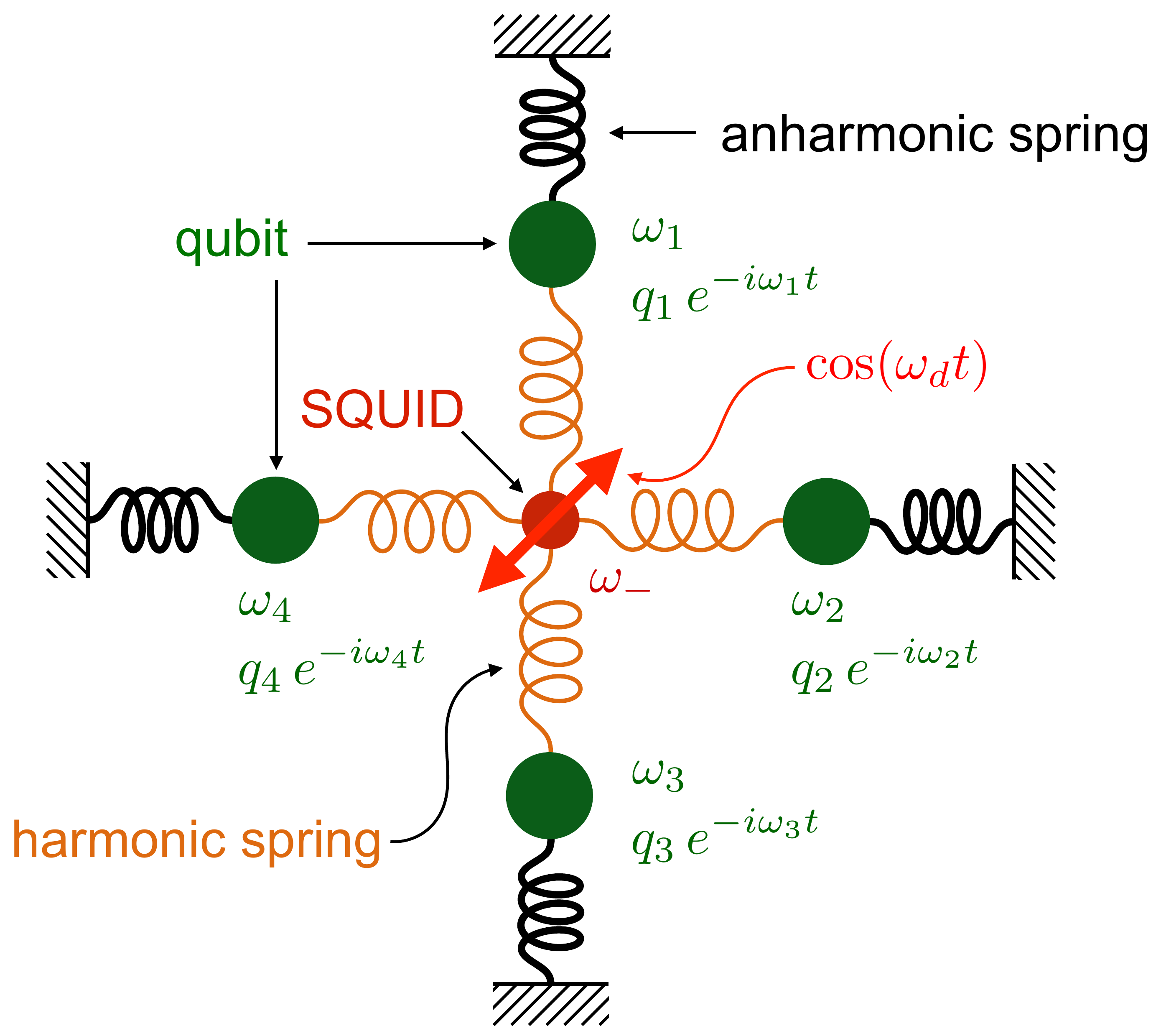}
\caption{\label{fig:analogy} Macroscopic analogy providing an intuitive explanation of our approach to generate the stabilizer interactions. The qubits are nonlinear oscillators so that one can think of them as masses (corresponding to the qubits' capacitances) attached to anharmonic springs (corresponding to the qubits Josephson energy). We here sketch four nonlinear oscillators representing four qubits as green spheres connected by black (anharmonic) springs to the support frame. Note that all four qubits have different transition frequencies so that all four spheres representing them here would have different masses and their springs different spring constants. For vanishing critical current the relevant SQUID modes $\varphi_{-;j}$, behave as harmonic oscillators and in our macroscopic analogy we can represent them by a mass (here sketched as a red sphere, corresponding to the capacitance $2 C_{J}$) that is attached via four springs (sketched in orange, corresponding to the inductors $L$) to the four masses representing the neighboring qubits. For vanishing critical current, the only potential energy of the modes $\varphi_{-;j}$ comes from their inductive coupling to the neighboring qubits, and hence the mass representing the SQUID mode (red sphere) has no spring connecting it to a support frame. We now imagine that we shake the mass representing the SQUID mode with a driving term that is proportional to the fourth power of the mass' position variable, $\varphi_{-;j}^4$. This corresponds to the oscillating flux bias $\propto \varphi_{ac} F(t)$. If we shake this mass at a frequency $\omega_{d}$ that is distinct from its eigenfrequency $\omega_{-;j}$ but equals the sum of the oscillation frequencies of all four neighboring masses (neighboring qubits), we drive a term proportional to $q_{1}^{\dagger}q_{2}^{\dagger}q_{3}^{\dagger}q_{4}^{\dagger}$, where $q_{j}^{\dag}$ creates an excitation in qubit $j$. If, on the other hand we shake the SQUID-mass at a frequency $\omega_{d} = \omega_{1}+\omega_{2}-\omega_{3}-\omega_{4}$ we drive a term proportional to $q_{1}^{\dagger}q_{2}^{\dagger}q_{3}q_{4}$. The selective driving of specific four-body terms, as we consider it here, of course relies on choosing the transition frequencies of all four qubits to be different. Our approach considers a superposition of various drive frequencies such that the sum of the triggered four-body terms adds up to the desired stabilizer interaction, see text.}
\end{figure}

Further intuition for our approach can be obtained from the macroscopic analogy explained in Fig.~\ref{fig:analogy}. In short, the plasma frequency of the coupling SQUIDs differs form all transition frequencies of the qubits they couple to as well as from all frequencies contained in the oscillating flux that is threaded through their loops. They thus decouple from the qubits but mediate the desired four-body interactions via the drives applied to them. Further details of this analogy are explained in the caption of Fig.~\ref{fig:analogy}. We now turn to present the quantitatively precise derivation.

\subsection{Elimination of the SQUID degrees of freedom} \label{sec:elim-squid}
For the parameters of our architecture, the oscillation frequencies of the SQUID modes $\varphi_{\pm;j}$ differ from the transition frequencies of the qubits and the frequencies of the time dependent fluxes threaded through the SQUID loops by an amount that exceeds their mutual couplings by more than an order of magnitude. These modes will thus remain in their ground states to very high precision. 
To eliminate them from the description and obtain an effective Hamiltonian for the qubits only, it is useful to consider the Schrieffer-Wolff transformation
\begin{align} \label{eq:SW-main}
  H &\to \tilde{H} = e^{S} H e^{-S} \quad \text{with}\\
  S & = \frac{i}{2 \hbar} \sum_{j=(1,1)}^{(N,M)} (\varphi_{q+;j} \pi_{+;j} + \varphi_{q-;j} \pi_{-;j}),\nn
\end{align}
which eliminates the qubit-SQUID interactions described in Eq.~(\ref{eq:qubit-squid-coupling}). The effective Hamiltonian for the qubits is then obtained by projecting the resulting Hamiltonian onto the ground states of the SQUID modes $\varphi_{\pm;j}$. The details of this derivation are discussed in appendix \ref{sec:eff-ham}.

For the transformed Hamiltonian $\tilde{H}$, see Eq.~(\ref{eq:SW-main}), we consider terms up to fourth order in $S$ to get the leading contributions of the terms which are in resonance with the oscillating part of $\varphi_{ext}(t)$, c.f. Eq.~(\ref{eq:drive}) .     
On the other hand, a truncation of the expansion is here well justified since both fields $\varphi_{q\pm;j}$ and $\pi_{\pm;j}$ have low enough amplitudes, see Eq.~(\ref{eq:SchriefferWolffamplitude}), so that higher order terms can safely be ignored. The transformed Hamiltonian can be written as
\begin{equation}\label{eq:transformed-ham}
 \tilde{H} = H_{0q} +H_{Iq} + H_S + H_{qS},
\end{equation}
where $H_{0q} = \sum_{j=(1,1)}^{(N,M)} H_{q;j}$ with $H_{q;j}$ as in Eq.~(\ref{eq:qubitham}) describes the individual qubits and  
\begin{equation}\label{eq:qq-ints}
H_{Iq} = - 2 E_{L} \sum_{n,m=1}^{N,M}  \varphi_{n,m} \varphi_{n,m+1} 
\end{equation}
interactions between neighboring qubits that are ineffective since their transition frequencies differ by an amount that exceeds the relevant coupling strength by more than an order of magnitude, see appendix \ref{sec:first-order-corrs-adiabat}.

The remaining two terms $H_{S}$ and $H_{qS}$ in Eq.~(\ref{eq:transformed-ham}) describe the SQUIDs and the residual couplings between the qubits and the SQUIDs, see Eqs.~(\ref{eq:ham-squids-1}) and (\ref{eq:nonlinear}) in appendix \ref{sec:eff-ham}. Since the oscillation frequencies of the SQUID modes are chosen such that they cannot be excited by the applied driving fields or via their coupling to the qubits, an effective Hamiltonian, describing only the qubits, can be obtained by projecting $H_{qS}$ onto the ground states of the SQUID modes,
\begin{equation} \label{eq:projectGS}
H_{qS} \to H_{qS}^{(0)} = \langle 0_{\text{SQUIDs}} | H_{qS} | 0_{\text{SQUIDs}} \rangle.
\end{equation}
As the SQUID modes remain in their ground states, their Hamiltonian $H_{S}$ may be discarded. The projected Hamiltonian $H_{qS}^{(0)}$
contains two-body interactions and four-body interactions between the qubits,
\begin{equation} \label{eq:projectGS1}
\begin{split}
H_{qS}^{(0)} & = - \frac{\chi \varphi_{ac} E_{J}}{8} \sum_{n,m=1}^{N,M} F(t) \varphi_{q-;n,m}^{2} \\
& + \frac{1}{4!} \frac{\chi \varphi_{ac} E_{J}}{16} \sum_{n,m=1}^{N,M}  F(t) \varphi_{q-;n,m}^{4}
\end{split}
\end{equation}
with $\chi = \langle 0_{-;n,m}| \cos (\varphi_{-;n,m}) | 0_{-;n,m} \rangle$, where $| 0_{-;n,m} \rangle$ is the ground state of the mode $\varphi_{-;n,m}$. 

We emphasize here that our approach requires nonlinear coupling circuits, such as the considered SQUIDs. The terms in the second line of Eq.~(\ref{eq:projectGS1}), which give rise to the desired four-body interactions, only emerge because the Hamiltonian of the coupling SQUIDs contains terms that are at least fourth power in $\varphi_{-;n,m}$, see  appendix \ref{sec:eff-ham}.

The oscillation frequencies contained in the external flux, i.e. in $\varphi_{ext}$, can be chosen such that selected terms in the second line of Eq.~(\ref{eq:projectGS1}) are enabled as they become resonant. For suitable amplitudes of the frequency components of $\varphi_{ext}$, these terms add up to the desired four-body stabilizer interactions whereas all other interactions contained in Eq.~(\ref{eq:projectGS1}) are for our choice of parameters strongly suppressed in a rotating wave approximation. 

Before discussing this point in more detail in the next section,
let us here emphasize  the versatility of our approach: 
With suitably chosen Fourier components of $\varphi_{ext}$ almost any two-, three- or four-body interaction can be generated. The only limitation here is that terms with the same rotation frequency can not be distinguished, e.g. one can not choose between generating $\sigma_{j}^{z} \sigma_{k}^{z}$ or $\sigma_{j}^{z}  \sigma_{l}^{z}$ for $j \ne k \ne l$ which are both non-rotating. On the other hand, straightforward modifications of the circuit allow to implement higher than four-body (e.g. five- and six-body) interactions.

\subsection{Effective Hamiltonian of one cell}
We now focus on the interactions between the four qubits in one cell only, see Fig.~\ref{fig:lattice-main}b. To simplify the notation we label these four qubits clockwise with indices 1,2,3 and 4, i.e. $(n,m) \to 1$, $(n+1,m) \to 2$, $(n+1,m+1) \to 3$ and $(n,m+1) \to 4$. 
We thus divide one term, say for $(n,m) = (n_{0}, m_{0})$, in the sum of Eq.~(\ref{eq:projectGS1}) into the following parts,
\begin{equation} \label{eq:split-Hamiltonian}
H_{\text{cell}}= \left. H_{qS}^{(0)}\right|_{\text{cell}} = H_1 + H_2 + H_3,
\end{equation}
where 
\begin{equation*}
 H_1 = \frac{\chi \varphi_{ac} E_{J}}{16} F(t) \varphi_1 \varphi_2 \varphi_3 \varphi_4,
\end{equation*}
contains the four-body interactions of the stabilizers,
\begin{equation*}
 H_2 = - \frac{\chi \varphi_{ac} E_{J}}{8}  F(t) (\varphi_1 + \varphi_4 - \varphi_2 - \varphi_3)^2,
\end{equation*}
contains all terms that are quadratic in the node phases $\varphi_{j}$ and 
\begin{equation*}
H_3 = -\chi \varphi_{ac} E_{J} F(t) \sum_{(i,j,k,l)} C_{ijkl} \varphi_i \varphi_j \varphi_k \varphi_l
\end{equation*}
contains all terms that are fourth-order in the node phases $\varphi_{j}$ except for the four-body interactions. These are terms where at least two of the indices $i,j,k$ and $l$ are the same as indicated by the notation $\sum_{(i,j,k,l)}$. 
Details of the derivation of the Hamiltonian (\ref{eq:split-Hamiltonian}) are given in appendix \ref{sec:rot-frame}, where we also state the exact form of the coefficients $C_{ijkl}$.

The Hamiltonian (\ref{eq:split-Hamiltonian}) can be written in second quantized form using
$\varphi_j = \overline{\varphi}_{j}(q_j + q_j^\dagger)$, where $q_j^{\dagger}$ ($q_j$) creates (annihiliates) a bosonic excitation at node $j$ with energy $ \hbar \omega_j = \sqrt{8 E_{C;j} (E_{Jq;j} + 4 E_{L})}$ and zero point fluctuation amplitude
$\overline{\varphi}_{j} = [2 E_{C;j} / (E_{Jq;j} + 4 E_{L})]^{1/4}$. Note that the effective Josephson energy $E_{Jq;j} + 4 E_{L}$ includes here contributions of the attached inductances, see appendix \ref{sec:2nd-quantized} for details.
Due to the qubit nonlinearities we can restrict the description to the two lowest energy levels of each node by replacing $q_j \to \sigma_j^{-}$ ($q_j^{\dag} \to \sigma_j^{+}$) and get for the four-body interactions,
\begin{equation} \label{eq:ham1-rot}
 H_1 = \frac{\chi \varphi_{ac} E_{J}}{16} F(t) \prod_{i=1}^{4}\overline{\varphi}_{i} \sum_{a,b,c,d \in \{+,-\}}\sigma_1^a\sigma_2^b\sigma_3^c\sigma_4^d,
\end{equation}
Here, the sum $\sum_{a,b,c,d \in \{+,-\}}$ runs over all 16 possible strings of $\sigma^{+}$ and $\sigma^{-}$ operators, which are listed in table \ref{tab:1} in appendix \ref{sec:rot-frame}.

The explicit form of the Hamiltonians $H_{2}$ in second quantized form is given in Eq.~(\ref{eq:two_body_rotating}). The terms contained in $H_2$ only lead to frequency shifts for the qubits and negligible two-body interactions. The frequency shifts can be compensated for by modified frequencies for the driving fields and the residual two-body interactions can be made two orders of magnitude smaller than the targeted four-body interactions, see Sec. \ref{sec:four-body-ints} and appendix \ref{sec:perturbations}. Moreover the terms contained in $H_3$ only lead to corrections that are at least an order of magnitude smaller than those of $H_2$ and can safely be discarded, see appendix \ref{sec:residual-2body-app}. Here, we therefore focus on the part $H_{1}$ which gives rise to the dominant terms in the form of four-body stabilizer interactions.

\subsection{Triggering four-body interactions}
\label{sec:four-body-ints}

In a rotating frame where each qubit rotates at its transition frequency,
$\sigma^+_j(t) = e^{i\omega_j t}\sigma^+_j$ [$\sigma^-_j(t) = e^{-i\omega_j t}\sigma^-_j$], the strings $\sigma_1^a\sigma_2^b\sigma_3^c\sigma_4^d$ that appear in Eq.~(\ref{eq:ham1-rot}) rotate at the frequencies $\omega_{a,b,c,d} = a \omega_1 + b \omega_2 + c \omega_3 + d \omega_4$. 
Hence, if we thread a magnetic flux, which oscillates at the frequency $\omega_{a,b,c,d}$, through the SQUID loop, the corresponding operator $\sigma_1^a\sigma_2^b\sigma_3^c\sigma_4^d$ acquires a pre-factor with oscillation frequency $\omega_{a,b,c,d}$, which renders the term non-oscillating and hence enables this particular four-body interaction.
Importantly, all qubits of a cell have different transition frequencies such that a drive with frequency $\omega_{a,b,c,d}$ only enables the particular interaction $\sigma_1^a\sigma_2^b\sigma_3^c\sigma_4^d$ whereas all other interactions are suppressed by their fast oscillating prefactors.

As the star and plaquette interactions $J_s\sigma_1^x\sigma_2^x\sigma_3^x\sigma_4^x$ and $J_p\sigma_1^y\sigma_2^y\sigma_3^y\sigma_4^y$ can be written as linear combinations of operator products $\sigma_1^a\sigma_2^b\sigma_3^c\sigma_4^d$, their generation requires magnetic fluxes that are a superposition 
of oscillations at the respective frequencies. 
To enable the specific four-body interactions of the Toric Code we thus consider,
\begin{equation}
F(t) =  \left\{
 \begin{array}{l}
 F_s = F_X + F_Y\\
 F_p = F_X - F_Y
 \end{array}
\right.
\end{equation}
where $F_X = f_{1,1,1,1} + f_{1,1,-1,-1} + f_{1,-1,1,-1} + f_{1,-1,-1,1}$
and $F_Y = f_{1,1,1,-1} + f_{1,1,-1,1} + f_{1,-1,1,1} + f_{-1,1,1,1}$
with $f_{a,b,c,d} = \cos(\omega_{a,b,c,d}t)$.
The driving field $F_s$ generates a star interaction $A_{s}$ and $F_p$ a plaquette interaction $B_{p}$.
Applying $F_s$ or $F_p$ to a cell, its Hamiltonian (\ref{eq:ham1-rot}) thus reads,
\begin{subequations}
\begin{align}
\left. H_{cell}^{(1)}(t) \right|_{F_{s}} & = - J_s\sigma_1^x\sigma_2^x\sigma_3^x\sigma_4^x  + r.t. \quad \text{for} \: \, F_s\label{subeq1}\\
\left. H_{cell}^{(1)}(t) \right|_{F_{p}} & = - J_p\sigma_1^y\sigma_2^y\sigma_3^y\sigma_4^y  + r.t. \quad \text{for} \: \, F_p\label{subeq2}
\end{align}
\end{subequations}
where $r.t.$ refers to the remaining fast rotating terms, and 
\begin{equation} \label{eq:intstrength}
J_s=J_p= \frac{\chi E_{J} \varphi_{ac}}{16 } \prod_{i=1}^{4}\overline{\varphi}_{i}.
\end{equation}
Hence star and plaquette interactions are realized with the same coupling circuit (shown in figure \ref{fig:lattice-main}b) and only differ by the relative sign of the $F_{X}$ and $F_{Y}$ contributions, i.e. a relative phase of $\pi$ in the respective Fourier components. Consequently, a driving field with $F_s$ ($F_p$) is applied to the coupling circuit of every orange (green) diamond in figure \ref{fig:lattice-main}a (see also figure \ref{fig:min-lattice}) and the resulting checkerboard pattern of driving fields generates the Toric Code Hamiltonian of equation (\ref{eq:toric-code-ham}).

Note that the signs and magnitudes of the interactions $J_{s}$ and $J_{p}$ can for each cell be tuned independently via the respective choices for the drive amplitudes $\varphi_{ac}$.

Besides the stabilizer interactions, there are further terms in the Hamiltonian (\ref{eq:transformed-ham}).
As we show in appendix \ref{sec:perturbations}, all these other interactions are strongly suppressed by their fast oscillating prefactors (the oscillation frequencies of the prefactors exceed the interaction strengths by more than an order of magnitude in all cases). 
Their effect can be estimated using time dependent perturbation theory. We find for our parameters that they lead to frequency shifts for the individual qubits, which can be compensated for by a modification of the frequencies of the driving field $\varphi_{ext}$. For our choice of qubit transition frequencies, where no nearest neighbors or next nearest neighbors of qubits have the same transition frequencies, the leading corrections to the Toric Code Hamiltonian of Eq.~(\ref{eq:toric-code-ham}), besides local frequency shifts, are two-body interactions between qubits that are further apart (i.e. neither nearest neighbors nor next neighrest neighbors). For suitable parameters as discussed in Sec.~\ref{sec:parameters}, these interactions are about two orders of magnitude weaker than the stabilizers and thus even weaker than dissipation processes in the device, see appendix \ref{sec:perturbations}.

In this context, we also note that the leading contribution to the perturbations of the Hamiltonian (\ref{eq:toric-code-ham}) is proportional to the charging energy $E_{Cq;j}$ of the employed qubits, c.f. appendix \ref{sec:perturbations}. Hence the considered transmon qubits are a well suited choice for our aims.

\subsection{Including dissipative processes}
\label{sec:full-dyn}
As we have shown, the dynamics of the architecture we envision can be described by the Hamiltonian (\ref{eq:toric-code-ham}) in a rotating frame with respect to $H_{0} = \sum_{j} \omega_{j} \sigma_{j}^{+} \sigma_{j}^{-}$, where we assume that all frequency shifts discussed in the previous section (see also appendix \ref{sec:perturbations}) have been absorbed into a redefinition of the $\omega_{j}$. In an experimentally realistic scenario, the qubits will however be affected by dissipation. The full dynamics of the circuit can therefore be described by the Markovian master equation,
\begin{equation} \label{eq:master}
\dot{\rho} = -i \hbar^{-1} [H,\rho] + \mathcal{D}_{r}[\rho] + \mathcal{D}_{d}[\rho]
\end{equation} 
where $\rho$ is the state of the qubit excitations, $H$ is as in Eq.~(\ref{eq:toric-code-ham}), $\mathcal{D}_{r}[\rho] = \frac{\kappa}{2} \sum_{j}(2 \sigma_{j}^{-} \rho \sigma_{j}^{+} - \sigma_{j}^{+} \sigma_{j}^{-} \rho - \rho \sigma_{j}^{+} \sigma_{j}^{-})$ describes relaxation at a rate $\kappa$ and $\mathcal{D}_{d}[\rho] = \kappa'\sum_{j}(\sigma_{j}^{z} \rho \sigma_{j}^{z} - \rho)$
pure dephasing at a rate $\kappa'$. Note that both, the relaxation and dephasing terms $\mathcal{D}_{r}[\rho]$ and $\mathcal{D}_{d}[\rho]$ are invariant under the transformation into the rotating frame. 

In the following we turn to discuss realistic experimental parameters that lead to an effective Toric Code Hamiltonian, the preparation and verification of topological order in experiments and protocols for probing the anyonic statistics of excitations in a realization of the Toric Code with our approach. In all these discussions, we fully take the relaxation and dephasing processes described by Eq.~(\ref{eq:master}) into account.

\section{Towards an experimental realisation}
\label{sec:parameters}
\begin{figure}
 \centering
 \includegraphics[width=0.6\columnwidth]{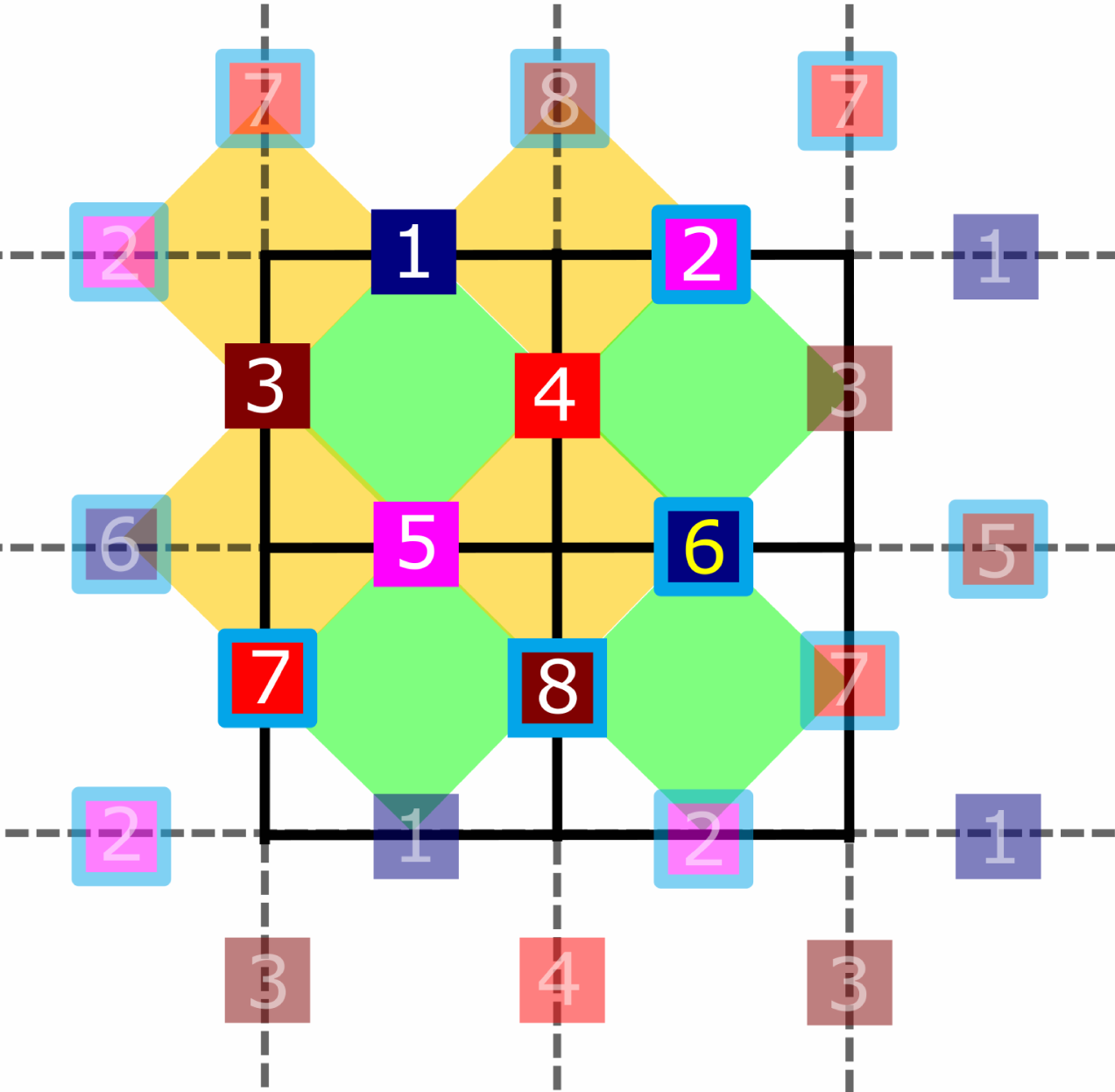}
 \caption{\label{fig:min-lattice} Lattice for minimal realization on a torus with 8 physical qubits including 4 star interactions (orange diamonds) and 4 plaquette interactions (green diamonds). The qubits are indicated by small colored squares, where the coloring encodes their transition frequencies and the numbering indicates the periodic boundary conditions. Here, 8 different transition frequencies are needed.}
\end{figure}

An experimental implementation of our approach is feasible with existing technology as it is based on transmon qubits which are the current state of the art in many laboratories. Moreover, periodic driving of a coupling element at frequencies comparable to the qubit transition frequencies is a well established technique \cite{Niskanen07,Roushan16}. The minimal lattice of a Toric Code with periodic boundary conditions in both lattice directions is the eight-qubit lattice sketched in Fig.~\ref{fig:min-lattice}. An important benefit of our rotating frame approach is its inherent insensitivity to disorder due to largely spaced qubit transition frequencies. The parameters we discuss in the sequel do thus not need to be realized exactly but only with moderate precision.

\subsection{Parameters}

A working example for this eight-qubit realization with experimentally realistic parameters is provided by the qubit transition frequencies given in table \ref{tab:transition-frequs}, which are realized in qubits with Josephson energies in the range $4.9 \, \text{GHz} \le E_{Jq}/h \le 94 \, \text{GHz}$ ($h$ is Planck's constant) and capacitances in the range $52 \, \text{fF} \le C_{q;j} \le 77 \, \text{fF}$. Note that the frequencies are chosen such that neighboring qubits are detuned from each other by several GHz whereas the transition frequencies of next nearest neighbor qubits differ by $\sim 100\,\text{MHz}$. A practical advantage of our approach with respect to frequency crowding is that it only requires a small set of continuous single frequency driving fields, whereas the frequency width of the pulses in digitized schemes grows inversely proportional to their durations.
\begin{table}[t]
\centering
\begin{tabular}{|c|c||c|c|}
\hline
qubit
& trans. frequency 
& qubit
& trans. frequency \\
& $\omega_{j}/2\pi$ & & $\omega_{j}/2\pi$ \\
\hline
1 & 13.8 GHz & 5 & 13.1 GHz\\
2 & 13.0 GHz & 6 & 13.7 GHz\\
3 & 3.5 GHz & 7 & 4.2 GHz \\
4 & 4.1 GHz & 8 & 3.4 GHz\\
\hline
\end{tabular}
\caption{\label{tab:transition-frequs} Possible transition frequencies for an eight-qubit Toric Code in natural frequency units. The qubit numbers correspond to the numbering chosen in Fig.~\ref{fig:min-lattice}. These frequencies are assumed to already include the frequency shifts calculated in appendix \ref{sec:perturbations}.}
\end{table}

For the coupling SQUIDs, the Josephson energy of each junction should be $E_{J}/h \approx 10 \, \text{GHz}$ and the associated capacitance $3.47 \, \textrm{fF}$. Since high transparency Josephson junctions typically have capacitances of 1$\,$fF per 50$\,$GHz of $E_{J}$, each SQUID is shunted by a capacitance of a few fF. For these parameters the modes $\varphi_{-;j}$ oscillate at $\omega_{-}/2\pi = 8.45\,\text{GHz}$, whereas the oscillation frequency of the modes $\varphi_{+;j}$ is much higher since $C_{g} \ll C_{J}$.

Furthermore the inductances coupling the qubits to the SQUIDs need to be chosen large enough so that the resulting coupling term remains weak compared to the qubit nonlinearities. This requires inductances of about $100 \, \text{nH}$ which can be built as superinductors in the form of Josephson junction arrays \cite{Masluk12,Bell12} or with high kinetic inductance superconducting nanowires \cite{Samkharadze16}. We note that our concept is compatible with nonlinearities Josephson junction arrays may have. The amplitude of the drive is $\varphi_{ac}=0.1\,$rad corresponding to an oscillating flux bias of amplitude $0.1\times\Phi_{0}$ at the SQUID.
These parameters give rise to a strength for star and plaquette interactions of $J_s=J_p= h \times 1 \,$MHz, which exceeds currently achievable dissipation rates of transmon qubits significantly \cite{Quintana14,Chow15,Bultink16,Chen16,OMalley16}. The strength of the stabilizer interactions can be enhanced further by increasing the Josephson energies of the SQUIDs $E_J$ and/or the amplitudes of the oscillating drives $\varphi_{ac}$.

For the above parameters, the total oscillation frequency, i.e. the sum of the frequencies of the qubit operators and the frequency of the external flux, is for any remaining two-body interaction term at least 10-30 times larger than the corresponding coupling strength. To leading order, these terms give rise to frequency shifts of the qubits  which can be compensated for by a modification of the drive frequencies. All interaction terms contained in higher order correction are negligible compared to the four-body interactions forming the stabilizers.
We note that the experimental verification of the topological phase is further facilitated by its robustness against local perturbations, such as residual two-body interactions or shifts of the qubit transition frequencies, c.f. \cite{Trebst07,Dusuel11}.

As an alternative to our approach, one could aim at implementing the Honeycomb model \cite{Kitaev06,You10}, which allows to obtain the Toric Code via a 4th order perturbative expansion in the limit where one of the spin-spin couplings strongly exceeds the two others \cite{Kitaev09}. This strategy would however require twice the number of qubits as our approach. For typical qubit-qubit couplings of $\sim$100\,MHz, it would moreover only lead to stabilizer interactions that are two orders of magnitude weaker than in our approach.

\subsection{Control and read-out}
In common with all two-dimensional lattices, the architecture we envision requires an increasing number of control lines for the qubits, SQUIDs and the read-out as one enhances the lattice size. Whereas solutions to this complexity challenge are already being developed, e.g. by placing the control lines on a second layer \cite{Barends14,Brecht16,Bejamin16}, a minimal realization of the Hamiltonian (\ref{eq:toric-code-ham}) requires only eight qubits that form four stars and four plaquettes, see Fig.~\ref{fig:min-lattice}. Moreover, these eight physical qubits are arranged on a 4 $\times$ 2 lattice so that every qubit is located on its boundary, facilitating the control access.

Controlling single qubit states and two qubit correlation measurements would require a charge line and a readout resonator for each qubit. Our scheme, where a single cell features four qubits with different transition frequencies, has the beneficial property that four readout resonators can be connected to a single transmission line and the qubit states read out using frequency division multiplexing \cite{Chen12,Jerger12,Schmitt14}. Such a joint readout is capable of measuring the state of all four qubits and qubit pairs even when their transition frequencies are separated by several GHz. An improvement over the readout through a common transmission line would be to employ a common Purcell filter \cite{Reed10,Jeffrey14} with two-modes, one mode placed at ~3.5 GHz and the second at ~12.5 GHz. Such scheme would not only allow a joint readout of four qubits but also protect them against the Purcell decay. Frequency division multiplexing can also be used for qubit excitation where a single charge line can be split or routed to four qubits in the cell. If a common transmission line is used for the readout the same line can also be used as a feedline for qubit state preparation \cite{Riste14}.

\subsection{Scalability}
\begin{figure}
 \centering
 \includegraphics[width=\columnwidth]{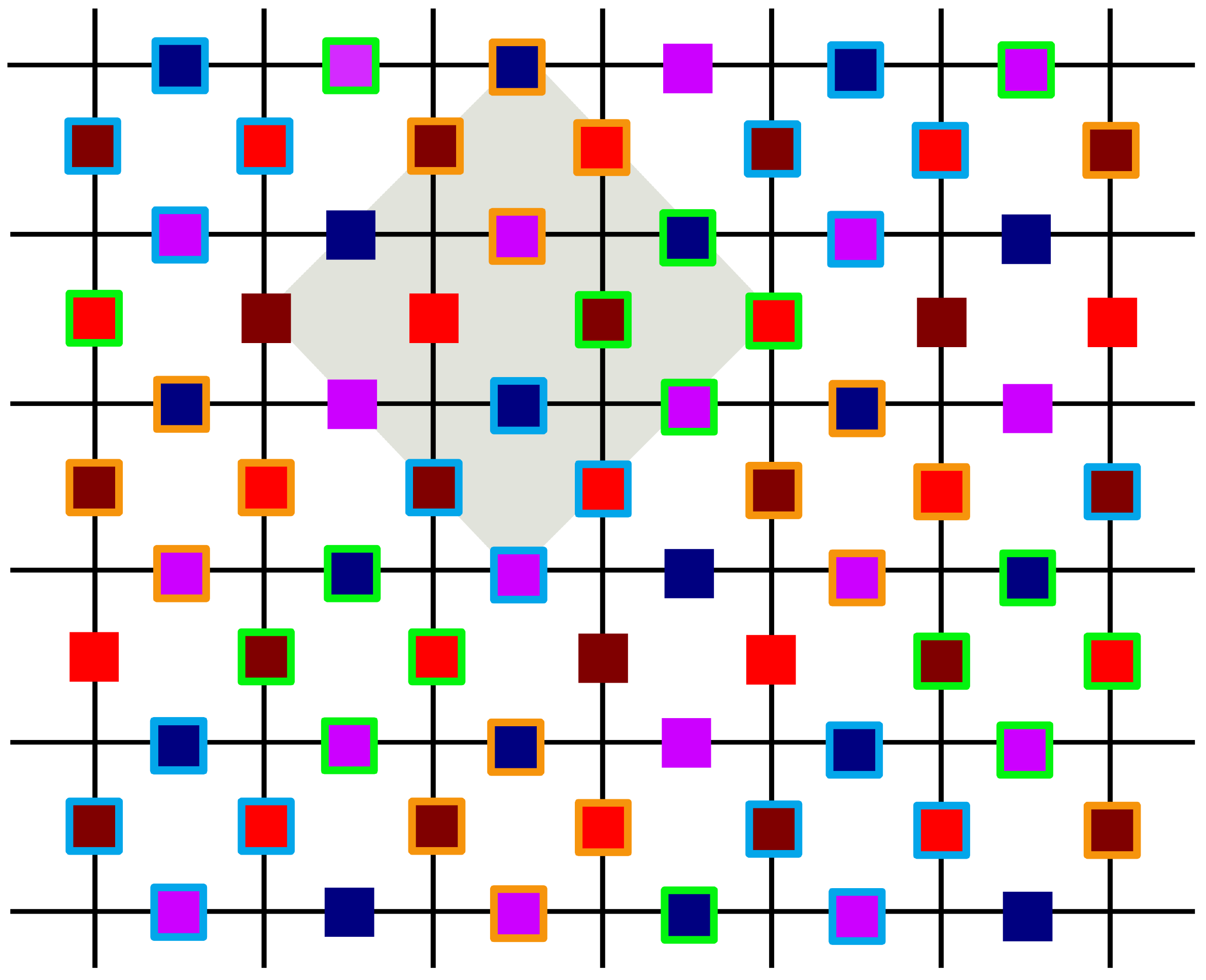}
 \caption{\label{fig:gen-lattice} Pattern of transition frequencies for an implementation of a large lattice. The qubits are indicated by small colored squares, where the coloring encodes their transition frequencies. Here four principal transition frequencies are needed. These differ by a few GHz and are indicated by the filling colors brown, blue, red and purple. To suppress two-body interactions between next nearest neighbor qubits, each pair of these needs to be detuned from each other by $\sim 100\,\text{MHz}$. The resulting frequency differences are indicated by the colored frames around the respective squares. The gray diamond highlights the frequency set which is periodically repeated in both lattice directions.}
\end{figure}

Importantly, our scheme is scalable as it also allows to realize the Toric Code on larger lattices. As already mentioned in sections \ref{sec:physical-lattice} and \ref{sec:four-body-ints}, the requirement for the choice of the qubit transition frequencies is that no nearest neighbor qubits and no next nearest neighbor qubits have the same transition frequencies. The resulting pattern of transition frequencies for a large lattice is sketched in Fig.~\ref{fig:gen-lattice}. It involves 16 qubit transition frequencies and can be realized with circuit parameters similar to the ones given above. These 16 qubit transition frequencies are periodically repeated in both lattice directions and therefore, in principle, sufficient for building any lattice size.

Larger and more realistic Toric Code realizations will obviously suffer from increased complexity. This could be drastically simplified when adding on-chip RF switches \cite{Pechal16} for routing microwave signals to different cells and using selective broadcasting technique which would also reduce the number of microwave devices such as signal generators and arbitrary waveform generators \cite{Asaad15}.

\section{Adiabatic preparation of topological order}
\label{sec:sweep}
On the surface of a torus, i.e. for a two dimensional lattice with periodic boundary conditions in both lattice directions,
the Hamiltonian (\ref{eq:toric-code-ham}) has four topologically ordered, degenerate eigenstates $|\psi_{a}\rangle$ ($a=1,2,3,4$), which are joint eigenstates of the stabilizers $A_{s}$ and $B_{p}$ with eigenvalue +1, see appendix \ref{sec:outcomes} for a possible choice of the $|\psi_{a}\rangle$ for an eight-qubit lattice.
A state in this manifold of topologically ordered states can either be prepared via measurements of the stabilizer operators or via an adiabatic sweep that starts from a product state \cite{Hamma08}. Measurements of stabilizers can in our architecture be performed via a joint dispersive readout of all qubits in a star or plaquette, see Sec.~\ref{sec:readout}. 

For the adiabatic sweep, one starts in our realisation with the oscillating driving fields turned off, $\varphi_{ac} = 0$ and  
the qubit transition frequencies blue-detuned by a detuning $\Delta$ with respect to their final values $\omega_j$. In the cryogenic environment of an experiment, this initializes the system in a product state, where all qubits are in their ground state, $\ket{\psi_{0}}=\prod_{j} \ket{0_j}$.
Then the amplitudes of the external oscillating fluxes, $\varphi_{ac}$, are gradually increased from zero to their final values $\varphi_{ac,f}$, i.e. $\varphi_{ac}(t) = \lambda (t) \, \varphi_{ac,f}$, where $\lambda (t=0) = 0$, $\lambda(T_{S}) = 1$, and $T_{S}$ is the duration of the sweep.
The qubit transition frequencies are decreased to $\omega_j$, i.e. $\omega'_j(t)-\omega_j = \Delta[1-\lambda(t)]>0$, with $\omega'_j(0)-\omega_j = \Delta$ and $\omega'_j(T_{S})-\omega_j = 0$. 
For this sweep, the Hamiltonian of the circuit can be written in the rotating frame as,
\begin{equation}
H(\lambda) = \sum_{j} (1-\lambda) \frac{\Delta}{2} \sigma^{z}_{j} - \lambda J_s \sum_s A_s - \lambda J_p \sum_p B_p,
\end{equation}
where $A_{s}$ and $B_{p}$ are as in Eq.~(\ref{eq:toric-code-ham}).
Thus the initial Hamiltonian is $H(\lambda=0)=\sum_{j} \Delta \sigma^{z}_{j}/2$ and the unique initial ground state is the vacuum with no qubit excitations (all spins down), i.e. $\ket{\psi_{0}}=\prod_{j} \ket{0_{j}} $. The final Hamiltonian is the targeted Toric Code Hamiltonian given in Eq.~(\ref{eq:toric-code-ham}).
Whereas the initial detuning ensures that the state $\ket{\psi_{0}}$ is also an eigenstate in the rotating frame for $J_{s} = J_{p} = 0$,
the sweep will drive the system adiabatically into the manifold of protected states, provided the parameters are tuned sufficiently slow.

\begin{figure}
 \centering
 \includegraphics[width=\columnwidth]{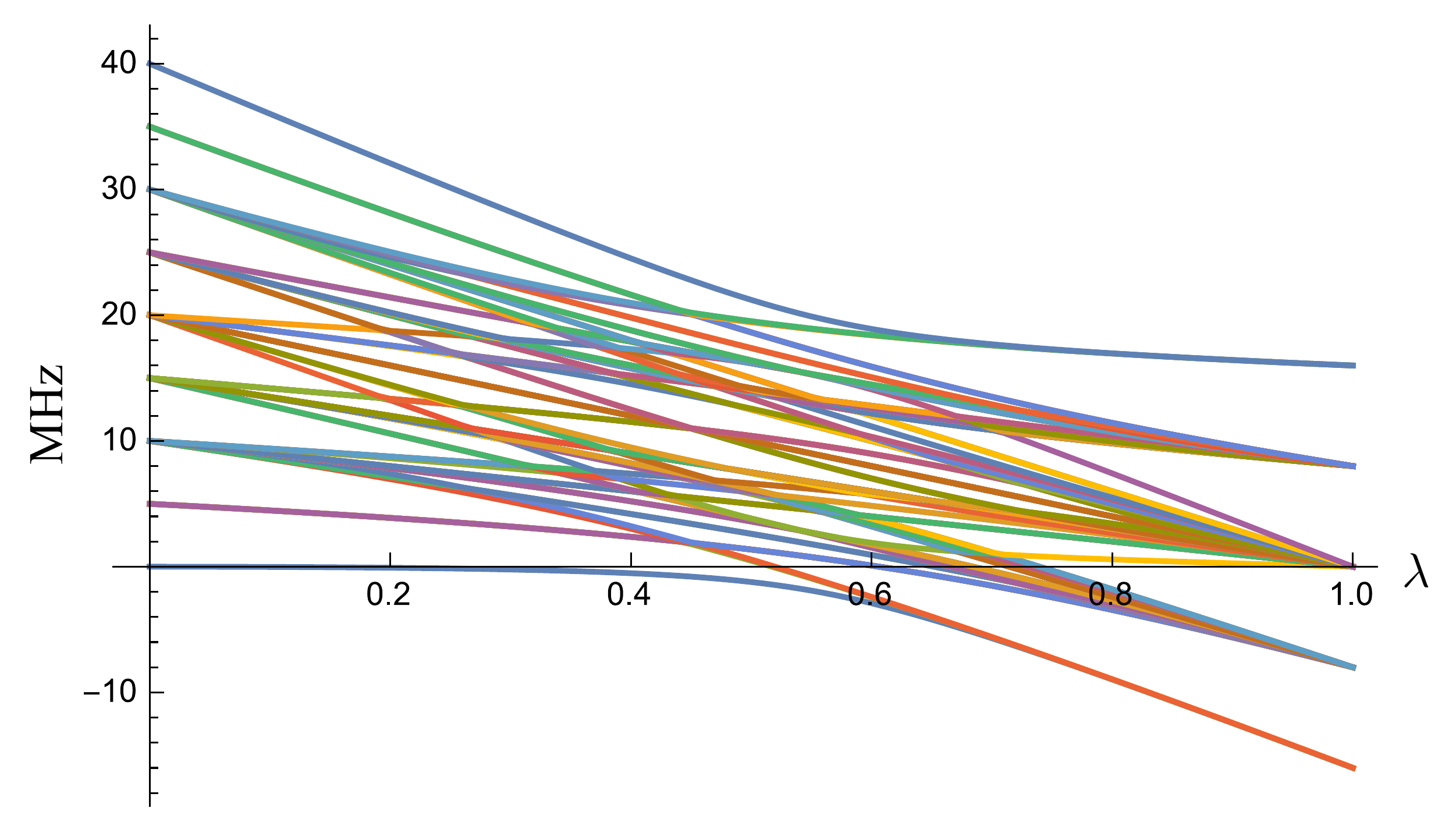}
\caption{\label{fig:adiabatic-spectrum} The energy eigenvalues of $H(\lambda)$ as a function of $\lambda$ for the eight-qubit lattice with $\Delta= 2 \pi \times 5$MHz and $J_s=J_p= 2 \pi \times 2$MHz.
The initial ground state is the unique vacuum $\ket{\psi_{0}}=\prod_{j=1}^8 \ket{0_j}$. At $\lambda \sim 0.6$ the ground state starts to become degenerate but the degenerate ground state manifold is always separated by a gap $\sim J_{s} \sim J_{p}$ from higher excited states.} 
\end{figure}

The spectrum of the minimal eight-qubit lattice, c.f. Fig.~\ref{fig:min-lattice}, is shown in Fig.~\ref{fig:adiabatic-spectrum} as a function of the tuning parameter $\lambda$. 
There is one unique ground state at $\lambda = 0$ which eventually becomes degenerate with three other states to form the degenerate four-dimensional manifold of topologically ordered ground states at $\lambda = 1$. In finite size lattices, there is however, for any value of $\lambda$, a finite energy separation of the order of $J_{s}$ or $J_{p}$ between the ground state and all states which do not end up in the topologically ordered manifold at $\lambda = 1$. We have numerically confirmed this for 8, 12 and 16 qubits. This spectral gap makes the sweep robust because it is not important which state in this manifold is prepared. All states in the manifold and any linear combination of them are equivalent. Therefore we use the probability of ending up in the topologically ordered manifold, $F = \sum_{a=1}^{4} \langle \psi_{a}| \rho(T_{S}) |\psi_{a}\rangle$, as a criterion for a successful preparation. Moreover, in the absence of dissipation one will end up in a pure state since a unitary evolution cannot change the purity of a state. To quantify how unavoidable dissipative processes in a real experiment would affect the preparation we also compute the purity of the final state $P= \text{Tr}(\rho^{2})$.

\begin{figure*}[t!]
 \centering
 \includegraphics[width=\textwidth]{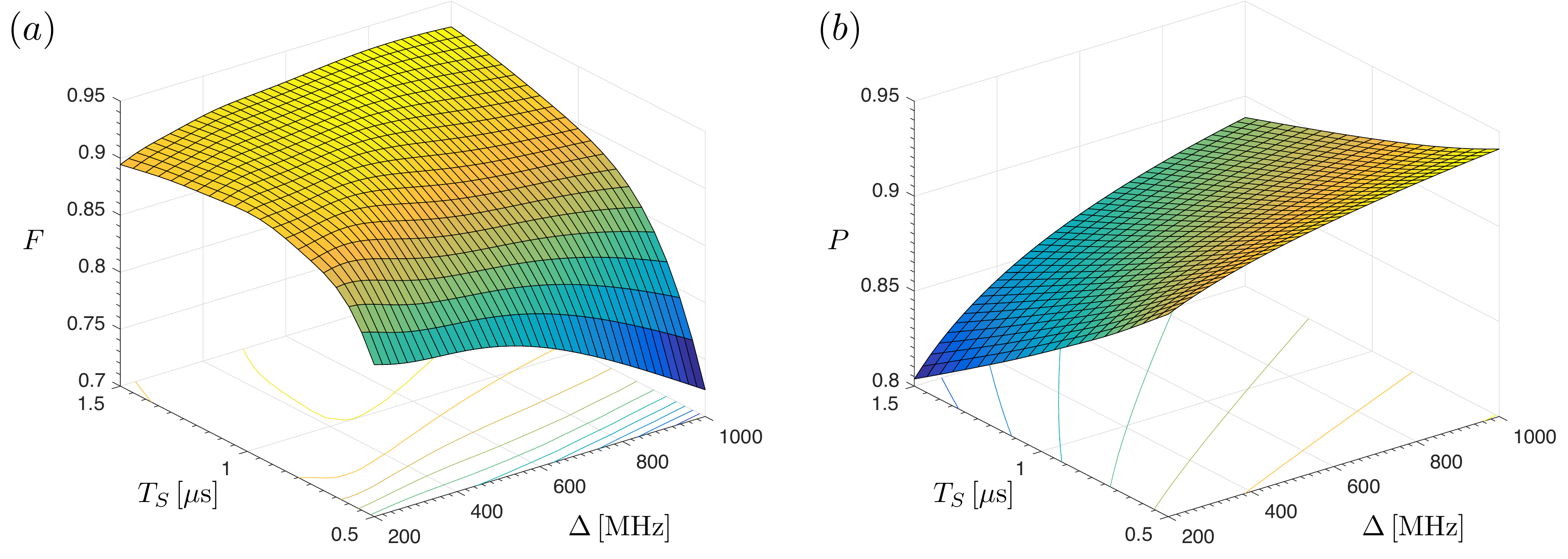}
\caption{\label{fig:adiabatic} Adiabatic preparation of an eight qubit lattice in the protected subspace.
{\bf (a)} Fidelity $F$ for being in the protected subspace and {\bf (b)} purity $P$ of the state. 
$J_s=J_p= h \times 2\,$MHz and coherence times $T_{1} = T_{2} = 50\,\mu$s, see Sec.~\ref{sec:full-dyn}.}
\end{figure*}

Figure \ref{fig:adiabatic}a shows the fidelity $F$ for being in the topologically ordered qubit subspace as a function of the sweep time $T_{S}$ and the initial detuning $\Delta$.
Here $\rho(t)$ is the actual state of the system in the rotating frame and the $|\psi_{a}\rangle$ ($a=1,\dots,4$) form the subspace of degenerate protected ground states. To test that the prepared state is pure, we also plot the purity $P$ as a function of $T_{S}$ and $\Delta$ in Fig.~\ref{fig:adiabatic}b. We choose $J_s=J_p= h \times 2 \,$MHz, see Sec.~\ref{sec:full-dyn}, relaxation and dephasing times $T_{1} = T_{2} = 50\,\mu$s, and a ramp profile $\lambda(t) = (1+\zeta ) (t/T_{S})^{6}/[1 + \zeta (t/T_{S})^{6}]$ with $\zeta = 100$. For these values, a protected state can successfully be prepared.

For larger lattices one may also follow the scheme proposed in \cite{Hamma08} which does not rely on a finite energy separation between the ground states and the higher excited states that do not become topologically ordered in the sweep. In our case, the scheme of \cite{Hamma08} proceeds as follows. All qubits, which are initially in their states $\ket{0_{j}}$, are first transformed into states $\ket{+_{j}}$ with $\sigma_{j}^{x}\ket{+_{j}} = +1 \ket{+_{j}}$ via a $\pi/2$-pulse. Since the resulting state $\prod_{j} \ket{+_{j}}$ is an eigenstate of all star stabilizers $A_{s}$, these may then be switched on abruptly together with a term $-B_{x} \sum_{j} \sigma_{j}^{x}$ which stabilizes the state  $\prod_{j} \ket{+_{j}}$.
Subsequently the plaquette interactions are turned on and the term $-B_{x} \sum_{j} \sigma_{j}^{x}$ is turned off in a sweep that is adiabatic since the relevant transition matrix elements of the instantaneous Hamiltonian always vanish \cite{Hamma08}.

Once they are prepared, it is of course desirable to stabilize the topologically ordered states as long as possible. Several approaches to this task have been put forward, see \cite{Kapit14,Bardyn16} and references therein, which are compatible with the rotating frame implementation we consider. Most approaches consider a shadow lattice of auxiliary qubits that are coupled to the primary lattice. In our architecture, such auxiliary systems may not be required as the coupling SQUIDs could be employed for this task.
 
\section{Measuring topological correlations}\label{sec:readout}
Our scheme allows to experimentally explore and verify topological order. To this end one prepares the lattice in one of the topologically ordered states $|\psi_{a}\rangle$ and measures the state of individual physical qubits as well as a correlation such as $\langle \prod_{l \in v} \sigma_{l}^{x}\rangle$ for a non-contractible path $v$ along one of the lattice directions, which is closed in a loop to implement the periodic boundary conditions. For the minimal eight qubit realization this is merely a measurement of a two-spin correlation, i.e. $\langle \sigma_{m}^{x}\sigma_{n}^{x}\rangle$.
Non-contractible loops of $\sigma^y$ operations, such as $\prod_{l \in v'} \sigma_{l}^{y}$, transform the topologically ordered states $|\psi_{a}\rangle$ into one another \cite{Kitaev09}. Here, these can be realized via strings of $\pi$-pulses applied to the qubits along the loop. 
After applying a string of $\pi$-pulses along one lattice direction, which transforms $|\psi_{a}\rangle$ into a different topologically correlated state $|\psi_{b}\rangle$, one again measures individual spins as well as the correlation $\langle \prod_{l \in v} \sigma_{l}^{x}\rangle$. Whereas the states of the individual spins should be the same for $|\psi_{a}\rangle$ and $|\psi_{b}\rangle$ because the topologically correlated states are locally indistinguishable, the results for the correlations should be distinct. 

The property of only being distinguishable via spin-spin correlations, but not by single-spin quantities, is not unique to topologically ordered states but also holds for Bell states. In contrast, a property that is unique to topologically ordered states is that local perturbations cannot transform one topologically ordered state $|\psi_{a}\rangle$ into another one $|\psi_{b}\rangle$. This crucial property can be experimentally tested with our approach. 

In the rotating frame we consider, single-spin perturbations on lattice site $j$ can be taken to read,
\begin{equation} \label{eq:loc-pert}
	H_{\text{pert},j}(t) = g(t) \sigma_j^\alpha \quad \text{for} \quad \alpha = + , - \: \: \text{or} \: \: z,
\end{equation} 
where oscillating prefactors of $\sigma_j^\alpha$ have been included into the time dependence of $g(t)$.
In this rotating frame, the excited states of the Toric Code Hamiltonian (\ref{eq:toric-code-ham}) are separated from the degenerate topologically ordered states by an energy $4 J$ (assuming $J_{s} = J_{p} = J$), irrespective of the lattice size. Therefore, the perturbations $H_{\text{pert},j}(t)$ cannot transform a topologically ordered state into a state outside the topologically ordered manifold provided $|g(t)| \ll 4J$. Yet because of the topological order, no local perturbation is capable of transforming a topologically ordered state  $|\psi_{a}\rangle$ into another one $|\psi_{b}\rangle$. As a consequence, provided their strength fulfills the condition $|g(t)| \ll 4J$, no local perturbation of the form (\ref{eq:loc-pert}) can affect a topologically ordered state $|\psi_{a}\rangle$. This robustness can be tested experimentally in our approach by verifying that the prepared topologically ordered states are insensitive to any local operation, provided it is weak enough. 

However, as indicated above, the topologically ordered states are in our rotating frame implementation not robust against local dissipation because the coupling to an environment can change the energy in the Toric Code lattice by an amount that exceeds $4J$.

This experimental verification of topological correlations can even be performed for the minimal lattice consisting of eight qubits, where it only requires measuring individual qubits and two-qubit correlations. An example of possible outcomes for these measurements are given in table \ref{tab:two-spi-corrs} and discussed in appendix \ref{sec:outcomes}.

In recent years several strategies for state tomography in superconducting circuits have been demonstrated experimentally \cite{Hofheinz09,Filipp09,DiCarlo10,Riste14}.
These either proceed by reading out the qubit excitations via an auxiliary phase-qubit or resonator. Both concepts are compatible with our circuit design.
In particular correlations between multiple qubits, including the stabilizers $A_{s}$ and $B_{p}$ can be measured in a joint dispersive read-out via a common coplanar waveguide resonator \cite{Filipp09,DiCarlo10,Riste14,Salathe15} or several readout resonators connected to a single transmission line using frequency multiplexing \cite{Chen12,Jerger12,Schmitt14}.

\section{Braiding and detection of anyons}\label{sec:anyons}

Our approach is also ideally suited for probing the fractional statistics of excitations in the Toric Code.
Once the lattice has been prepared in a topologically ordered state, for example via the sweep described in Sec.~\ref{sec:sweep}, the anyonic character of the excitations can be verified in an experiment using single qubit rotations and the measurement of one stabilizer only. We here discuss a protocol that has originally been proposed for cold atoms \cite{Han07} but is very well suited for superconducting circuits.

For the prepared topologically ordered state $|\psi\rangle$, the protocol considers applying a single qubit rotation $\sigma^{y}$ to one of the qubits.
This operation, which in our approach can be implemented via a microwave pulse in a charge line to the qubit, creates a pair of so called $e$-particles \cite{Kitaev09} on the neighboring vertices and puts the system into a state $|\psi,e\rangle$. Alternatively, half a $\sigma^{y}$-rotation creates the state $(|\psi\rangle + |\psi,e\rangle)/\sqrt{2}$. Similarly, with a $\sigma^{x}$-rotation on another qubit, one creates a pair of $m$-particles. Then one can move one of these $m$-particles around one of the $e$-particles via successive $\sigma^{x}$-rotations on the qubits along the chosen path and fuse both $m$-particles again.
After this sequence, the system is in the state $(|\psi\rangle - |\psi,e\rangle)/\sqrt{2}$ due to the fractional phase $\pi$ acquired from braiding the
$e$- and $m$-particles.
This phase flip can for instance be detected with another $\sqrt{\sigma^{y}}$-operation that maps $(|\psi\rangle - |\psi,e\rangle)/\sqrt{2} \to |\psi\rangle$ and
$(|\psi\rangle + |\psi,e\rangle)/\sqrt{2} \to |\psi,e\rangle$ and a subsequent measurement of the corresponding stabilizer $A_{s}$. We note that in our approach all single qubit operations can be implemented via pulses applied through charge lines and the stabilizer measurement can be realized via the technique described in Sec.~\ref{sec:readout}.

\section{Conclusion and Outlook}

We have introduced a scheme and architecture for a quantum simulator of topological order that implements the central model of this class of quantum many-body systems, the Toric Code, on a two dimensional superconducting circuit lattice. Our approach is based on a scheme that allows to realize a large variety of many-body interactions by coupling multiple qubits via dc SQUIDs that are driven by a suitably oscillating flux bias. The advanced status of experimental technology for superconducting circuits makes available an ample toolbox for exploring the physics of topological order with our scheme.

All physical qubits and coupling SQUIDs can be individually controlled with high precision, topologically ordered states can be prepared via an adiabatic ramp of the stabilizer interactions, multi-qubit correlations including stabilizers and correlations along non-contractible loops can be measured and anyons can be braided and their fractional statistics can be verified. Moreover, superconducting circuit technology allows for realizing periodic boundary conditions, a feature that not easily available in other platforms but crucial for exploring topological order.
We thus expect our work to open up a realistic path towards an experimental investigations of this intriguing area of quantum many-body physics that at the same time will enable access for detailed measurements.

Our scheme is highly versatile and not restricted to implementations of the Toric Code only. Several directions for generalizations appear very intriguing at this stage. 

(i) By tuning the time-independent flux bias of the coupling SQUIDs away from the operating point for the Toric Code, one can generate additional two-body interactions of the form $\sigma^{z}_{j}\sigma^{z}_{l}$ between neighboring qubits. These can become stronger than the stabilizer interactions. The ability to tune the transition frequencies of the qubits makes our approach a very versatile quantum simulator for exploring quantum phase transitions between topologically ordered and other phases of two-dimensional spin lattices. 

(ii) In our concept, the many-body interactions mediated by SQUIDs are not limited to 4-body interactions. By connecting one port of a coupling SQUID to $k_{l}$ qubits and the other to $k_{r}$ qubits, $k$-body interactions for $k=k_{l}+k_{r}$ can be implemented in $k$th order of the Schrieffer-Wolff expansion (\ref{eq:SW-main}) together with a suitable choice of frequencies for the flux bias $\varphi_{ext}$. Thus 3-body interactions, for example, that are an order of magnitude larger than the stabilizer interactions of equation (\ref{eq:intstrength}) can be implemented. For the parameters considered in section \ref{sec:parameters}, also 6-body interactions that are stronger than dissipation would be feasible and even interactions among a larger number of qubits are within reach with some improvement of circuit parameters. These capabilities of our scheme open the door to realizations of higher dimensional Toric Codes \cite{Dennis02,Bombin13} and other topological codes such as color codes \cite{Bombin06}. A useful property here is that the implementation of the sum of a star and plaquette interaction in one cell would only require half the frequency components for the driving fields.  

(iii) If two qubits that are connected to the same port of a coupling SQUID are identified with one lattice site of the simulated model, our approach can straight forwardly be generalized to the analogue quantum simulation of non-Abelian lattice gauge theories, where the engineered 4-body interactions play the role of the coupling matrix between the color degrees of freedom of the adjacent lattice sites.

(iv) Considering regimes, where the degrees of freedom of the coupling SQUID can no longer be adiabatically eliminated, one can modify our approach towards quantum simulations of dynamical gauge fields theories, where the coupling SQUID would represent the dynamical degrees of freedom of the gauge field.

\section{Acknowledgements}
M.~S. and M.~J.~H. are grateful for discussions with Martin Leib at an early stage of this project.
A.~P. and A.~W. acknowledge support by ETH Z\"urich.
M.~J.~H. was supported by the EPSRC under grant No. EP/N009428/1.
M.~S. and M.~J.~H. were supported in part by the National Science Foundation under Grant No. NSF PHY11-25915 for a visit to the program ``Many-Body Physics with Light'' at KITP Santa Barbara where a part of this work has been done.

\appendix

\section{Lagrangian and Hamiltonian of entire circuit} \label{sec:lagrangian}
In this appendix we present a detailed derivation of the Hamiltonian in Eq.~(\ref{eq:ham-full}) that describes the considered circuit lattice.
We start with the full Lagrangian describing the lattice depicted in Fig.~\ref{fig:lattice-main}, which reads,
\begin{widetext}
\begin{align} \label{eq:lagrangian}
\mathcal{L} & = \sum_{n,m=1}^{N,M} \left[\phi_{0}^{2} \frac{C_{q;n,m}}{2} \dot{\varphi}_{n,m}^{2} + E_{Jq;n,m} \cos(\varphi_{n,m}) \right]  \\
& - \sum_{n,m=1}^{M,N} \frac{\phi_0^2}{2 L}\left[ (\varphi_{a;n,m}-\varphi_{n,m})^2 + (\varphi_{a;n,m}-\varphi_{n,m+1})^2 + (\varphi_{b;n,m}-\varphi_{n+1,m})^2 + (\varphi_{b;n,m}-\varphi_{n+1,m+1})^2 \right] \nn \\
& + \sum_{n,m=1}^{M,N} \phi_{0}^{2} \left[\frac{C_{g}}{2} \dot{\varphi}_{a;n,m}^{2} + \frac{C_{g}}{2} \dot{\varphi}_{b;n,m}^{2} + C_{J} (\dot{\varphi}_{a;n,m}-\dot{\varphi}_{b;n,m})^{2}\right] + \sum_{n,m=1}^{M,N} 2E_J\cos(\frac{\varphi_{ext}}{2}) \cos(\varphi_{a;n,m}-\varphi_{b;n,m}). \nonumber
\end{align}
\end{widetext}
As explained in the main text, $\varphi_{n,m}$ is the phase at node $(n,m)$, $C_{q;n,m}$ is the capacitance and $E_{Jq;n,m}$ the Josephson energy of the qubit at that node.
$\varphi_{a;n,m}$ and $\varphi_{b;n,m}$ are the phases at the auxiliary nodes next to the SQUID located between nodes $(n,m)$, $(n,m+1)$, $(n+1,m)$, and $(n+1,m+1)$. 
The $C_{g}$ are the capacitances of these auxiliary nodes with respect to ground and $C_{J}$ is the capacitance of each junction in the SQUID including a shunt capacitance parallel to the junctions.
$E_{J}$ is the Josephson energy of each junction in the SQUID.

As the Josephson junctions of the dc-SQUIDs only couple to the difference between the phases of the two adjacent auxiliary nodes, we define the modes,
\begin{equation}
\varphi_{\pm;n,m} = \varphi_{a;n,m} \pm \varphi_{b;n,m}, 
\end{equation}
and describe the dc-SQUIDs in terms of $\varphi_{+;n,m}$ and $\varphi_{-;n,m}$ from now on.
We now turn to derive the Hamiltonian associated to the Lagrangian in Eq.~(\ref{eq:lagrangian}).

As there is neither capacitive coupling between the qubits in the lattice nor between the qubits and the coupling SQUIDs, the kinetic energy can be decomposed into the contribution of individual qubits and that of the individual coupling SQUIDs,
\begin{equation}
 \mathcal{K} = \sum_{n,m=1}^{N,M} \frac{\phi_0^2}{2} \left( C_{q;n,m} \dot{\varphi}_{n,m}^2 + \dot{\vec{\Phi}}_{s;n,m}^{T} C_{s} \dot{\vec{\Phi}}_{s;n,m}\right),
\end{equation}
where $\vec{\Phi}_{s;n,m} = (\varphi_{+;n,m}, \varphi_{-;n,m})^T$ and 
\begin{equation}
 C_s =\begin{pmatrix}
  2 C_J + \frac{C_g}{2}   & 0 \\
   0  & \frac{C_g}{2}
 \end{pmatrix}
\end{equation}
is the capacitive matrix associated with one SQUID.

The Hamiltonian of the lattice is obtained in the standard way via a Legendre transform of the Lagrangian $\mathcal{L}$ in Eq.~(\ref{eq:lagrangian}) by defining the qubit and SQUID momenta,
\begin{equation}
\pi_{n,m} = \frac{\partial \mathcal L}{\partial \dot{\varphi}_{n,m}}, \quad \text{and} \quad
\pi_{\pm;n,m} = \frac{\partial \mathcal L}{\partial \dot{\varphi}_{\pm;n,m}}.
\end{equation}
The kinetic energy thus reads,
\begin{equation}\label{eq:kin-en1}
\mathcal{K}  = \sum_{n,m=1}^{N,M} \left[ \frac{\pi_{n,m}^2}{2\hbar\phi_0^2 C_{q;n,m}} + \frac{\pi_{-;n,m}^{2}}{4 \hbar\phi_0^2 C_J} + \frac{\pi_{+;n,m}^{2}}{\hbar\phi_0^2 C_g} \right]
\end{equation}
where we have used that $C_g \ll C_J$ in inverting $C_s$.
%
%
We thus arrive at the Hamiltonian,
\begin{widetext}
\begin{align} 
\mathcal{H} & = \sum_{n,m=1}^{N,M} \left[4 \frac{E_{Cq;n,m}}{\hbar^{2}} \pi_{n,m}^{2} + 4 E_{L} \varphi_{n,m}^2 - E_{Jq;n,m} \cos(\varphi_{n,m}) \right] \label{eq:Hamiltonian}\\
&+\sum_{n,m=1}^{N,M} \left[4 \frac{E_{C+}}{\hbar^{2}}\pi_{+;n,m}^{2} + 4 \frac{E_{C-}}{\hbar^{2}} \pi_{-;n,m}^{2} + E_{L} \varphi_{+;n,m}^2 + E_{L} \varphi_{-;n,m}^2  - 2E_J\cos\left(\frac{\varphi_{ext}}{2}\right) \cos(\varphi_{-;n,m}) \right] \nn \\
& - \sum_{n,m=1}^{M,N} E_L \left[ \varphi_{q+;n,m}\varphi_{+;n,m}+\varphi_{q-;n,m}\varphi_{-;n,m} \right], \nn
\end{align}
\end{widetext}
where we have introduced the notation
\begin{equation}
\varphi_{q\pm;n,m} = \varphi_{n,m}+\varphi_{n,m+1} \pm \varphi_{n+1,m} \pm \varphi_{n+1,m+1}.
\end{equation}
$E_L = \phi_0^2/(2 L)$ is the inductive energy associated with an inductor $L$ and 
the charging energies of the qubits and SQUID modes are 
\begin{equation*}
E_{Cq;n,m} = \frac{e^2}{2 C_{q;n,m}}, \quad E_{C+} = \frac{e^2}{C_g} \quad \text{and} \quad E_{C-} = \frac{e^2}{4 C_J}, 
\end{equation*}
where $e$ is the elementary charge.
The first line in Eq.~(\ref{eq:Hamiltonian}) describes the qubits,
the second line the SQUIDs and the third line the interactions between qubits and SQUIDs via the inductances $L$. 

The Hamiltonian (\ref{eq:Hamiltonian}) is quantized in the standard way by imposing the canonical commutation relations
\begin{equation} \label{eq:canonical-commutations}
\left[\varphi_{j},\pi_{l}\right] = i \hbar \delta_{j,l} \quad \text{and} \quad  \left[\varphi_{\pm;j},\pi_{\pm;l}\right] = i \hbar \delta_{j,l}.
\end{equation}
The external flux applied through the SQUID loops and hence the associated phase is composed of a constant dc-part and a time dependent ac-part as given in Eq.~(\ref{eq:drive}). We choose $\varphi_{dc} = \pi$ and $\varphi_{ac} \ll \varphi_{dc} = \pi$, and obtain to leading order in $\varphi_{ac}$,
\begin{equation}
 \cos\left(\frac{\varphi_{ext}}{2}\right) \approx - \frac{\varphi_{ac}}{2} F(t).
\end{equation}

\section{Effective time-dependent Hamiltonian}  \label{sec:eff-ham}
In this appendix we present details of the derivation of the effective Hamiltonian for the qubits only, which will be dominated by the four-body interactions corresponding to the stabilizer operators in the Toric Code Hamiltonian, see Eq.~(\ref{eq:toric-code-ham}).

As the modes $\varphi_{\pm;n,m}$ of the SQUIDs oscillate at frequencies that strongly differ from the transition frequencies of the qubits and the oscillation frequencies of the applied flux biases, they remain in their ground states to a high degree of precision and can be adiabatically eliminated from the description. To this end we apply the Schrieffer-Wolf transformation introduced in Eq.~(\ref{eq:SW-main}) with the generator
\begin{equation} \label{eq:SW-app}
  S = \frac{i}{2 \hbar} \sum_{n,m=1}^{N,M} (\varphi_{q+;n,m} \pi_{+;n,m} + \varphi_{q-;n,m} \pi_{-;n,m}),
\end{equation}
and expand the transformed Hamiltonian up to fourth order in $S$,
\begin{widetext}
\begin{equation} \label{eq:SchriefferWolffexpansion}
 \tilde{H} = e^{S} \mathcal{H} e^{-S}
           = \mathcal{H} + [S,\mathcal{H}] + \frac{1}{2!}[S,[S,\mathcal{H}]] + \frac{1}{3!} [S,[S,[S,\mathcal{H}]]] + \frac{1}{4!} [S,[S,[S,[S,\mathcal{H}]]]] + \dots
\end{equation}
\end{widetext}
Using the canonical commutation relations (\ref{eq:canonical-commutations}), 
the commutators in the expansion (\ref{eq:SchriefferWolffexpansion}) read,
\begin{align*}
& [S,\mathcal{H}]  = \\
& \quad \quad = \sum_{n,m=1}^{N,M} E_L \varphi_{q+;n,m} \varphi_{+;n,m} + E_L \varphi_{q-;n,m} \varphi_{-;n,m}\\
& \quad \quad -4 \sum_{n,m=1}^{N,M} E_{Cq;n,m} \pi_{n,m} \left(\pi_{S-;n,m} + \pi_{S+;n,m} \right)\nn\\
& \quad \quad -\sum_{n,m=1}^{N,M} \frac{E_L}{2} (\varphi_{q+;n,m}^2 + \varphi_{q-;n,m}^2)\nn\\
& \quad \quad + \sum_{n,m=1}^{N,M} E_J \cos\left(\frac{\varphi_{ext}}{2}\right) \sin(\varphi_{-;n,m}) \varphi_{q-;n,m}\nn\\
& [S,[S,\mathcal{H}]] =\\
& \quad \quad = \sum_{n,m=1}^{N,M} \frac{E_L}{2} (\varphi_{q+;n,m}^2 + \varphi_{q-;n,m}^2) \\
& \quad \quad + 2 \sum_{n,m=1}^{N,M} E_{Cq;n,m} \left(\pi_{S-;n,m} + \pi_{S+;n,m} \right)^{2}\nn\\
& \quad \quad +\sum_{n,m=1}^{N,M} \frac{E_J}{2}\cos\left(\frac{\varphi_{ext}}{2}\right) \cos(\varphi_{-;n,m}) \varphi_{q-;n,m}^2\nn\\
& [S,[S,[S,\mathcal{H}]]] = \\
& \quad \quad = - \sum_{n,m=1}^{N,M} \frac{E_J}{4} \cos\left(\frac{\varphi_{ext}}{2}\right) \sin(\varphi_{-;n,m}) \varphi_{q-;n,m}^3 \\
& [S,[S,[S,[S,\mathcal{H}]]]] = \\
& \quad \quad =  -\sum_{n,m=1}^{N,M} \frac{E_J}{8}\cos\left(\frac{\varphi_{ext}}{2}\right) \cos(\varphi_{-;n,m}) \varphi_{q-;n,m}^4 
\end{align*}
where we have here introduced the abbreviations,
\begin{equation}
\pi_{S\pm;n,m} = \pi_{\pm;n,m}+\pi_{\pm;n,m-1}\pm\pi_{\pm;n-1,m}\pm\pi_{\pm;n-1,m-1}.
\end{equation}
We note that the terms $[S,[S,[S,\mathcal{H}]]]$ and $[S,[S,[S,[S,\mathcal{H}]]]]$ do not contain any contributions from the modes $\varphi_{+;n,m}$ which are just harmonic oscillators. They would vanish for an entirely harmonic coupling circuit and only the Josephson terms of the coupling SQUIDs that are nonlinear in $\varphi_{-;n,m}$ lead to contributions.

Truncating the expansion (\ref{eq:SchriefferWolffexpansion}) at fourth order is a good approximation since the amplitudes of the terms $\varphi_{n,m} \pi_{\pm;n,m}/\hbar$ are much smaller than unity, see Eq.~(\ref{eq:SchriefferWolffamplitude}) for details. Moreover, as we explain in the sequel, the frequencies contained in the flux bias $\phi_{ext}$ that is applied to the SQUIDs can be chosen such that interactions contained in the fourth order terms $[S,[S,[S,[S,\mathcal{H}]]]]$ become the dominant terms of the expansion (\ref{eq:SchriefferWolffexpansion}).

As stated in Eq.~(\ref{eq:transformed-ham}) of Sec.~\ref{sec:elim-squid}, the transformed Hamiltonian $\tilde{H}$ can be written as a sum of a part $H_{0q}$ describing the individual qubits, a part $H_{Iq}$ describing pure qubit-qubit interactions, a part $H_{S}$ describing the coupling SQUIDs and a part $H_{qS}$ describing the residual interactions between qubits and SQUIDs,
\begin{equation} \label{eq:transformed-ham-app}
 \tilde{H} = H_{0q} +H_{Iq} + H_S + H_{qS}.
\end{equation}
The Hamiltonian of the individual qubits, $H_{0q}$, and the qubit-qubit interactions, $H_{Iq}$, are given in Eqs.~(\ref{eq:transformed-ham}) and (\ref{eq:qq-ints}). We here repeat them for completeness,
\begin{align}
H_{0q} = & \sum_{n,m=1}^{N,M} H_{q;n,m} \quad \text{with}\\
H_{q;n,m} = & 4 \frac{E_{Cq;n,m}}{\hbar^{2}} \pi_{n,m}^{2} + 2 E_{L} \varphi_{n,m}^2 - E_{Jq;n,m} \cos(\varphi_{n,m}) \nn
\end{align}
and
\begin{equation} \label{eq:qubit-int}
H_{Iq} =  - 2 E_{L} \sum_{n,m=1}^{N,M} \varphi_{n,m} \varphi_{n,m+1} 
\end{equation}
The term describing the SQUIDs reads,
\begin{align} \label{eq:ham-squids-1}
H_{S}  & = \sum_{n,m=1}^{M,N} \left[ 4 \frac{\tilde{E}_{C+;n,m}}{\hbar^{2}}\pi_{+;n,m}^{2} + E_{L} \varphi_{+;n,m}^2 \right]\\
& + \sum_{n,m=1}^{M,N} \left[ 4 \frac{\tilde{E}_{C-;n,m}}{\hbar^{2}} \pi_{-;n,m}^{2} + E_{L} \varphi_{-;n,m}^2\right] \nn \\
&  - 2 E_J \cos\left(\frac{\varphi_{ext}}{2}\right) \sum_{n,m=1}^{M,N} \cos(\varphi_{-;n,m}), \nn 
\end{align}
where $\tilde{E}_{C+;n,m}$ and $\tilde{E}_{C-;n,m}$ are the renormalized charging energies of the $+$ and $-$ modes,
$\tilde{E}_{C\pm;n,m} =  E_{C\pm;n,m} + (E_{q;n,m} + E_{q;n,m+1} + E_{q;n+1,m} + E_{q;n+1,m+1})/4$.
The qubit-SQUID interactions read,
\begin{align} \label{eq:nonlinear}
  H_{qS}  =  &-4 \sum_{n,m=1}^{N,M} \frac{E_{Cq;n,m}}{\hbar^{2}} \pi_{n,m} \pi_{S-;n,m}\\
             &+ E_J\sum_{n,m=1}^{N,M}  \cos\left(\frac{\varphi_{ext}}{2}\right) \sin(\varphi_{-;n,m}) \varphi_{q-;n,m}\nn \\
             &+\frac{1}{2!}\frac{E_J}{2}\sum_{n,m=1}^{N,M} \cos\left(\frac{\varphi_{ext}}{2}\right) \cos(\varphi_{-;n,m}) \varphi_{q-;n,m}^2 \nn \\
             & - \frac{1}{3!}\frac{E_J}{4} \sum_{n,m=1}^{N,M}  \cos\left(\frac{\varphi_{ext}}{2}\right) \sin(\varphi_{-;n,m}) \varphi_{q-;n,m}^3 \nn \\
             & - \frac{1}{4!}\frac{E_J}{8} \sum_{n,m=1}^{N,M} \cos\left(\frac{\varphi_{ext}}{2}\right) \cos(\varphi_{-;n,m}) \varphi_{q-;n,m}^4 ,\nn 
\end{align}
where again we have neglected terms $\propto \pi_{n,m} \pi_{S+;n,m}$ because $|\omega_{+} - \omega_{n,m}| \gg |\omega_{-} - \omega_{n,m}|$, see Eq.~(\ref{eq:squid-frequs}), which makes these terms highly off-resonant so that we only keep the dominant contributions $\propto \pi_{n,m} \pi_{S-;n,m}$.

\subsection{Second-quantized form} \label{sec:2nd-quantized}
The free Hamiltonian of the qubits can be written in second-quantized form by defining the ladder operators $q_{i}$ and $q_{i}^{\dag}$ via
\begin{align} \label{eq:second-quant-qubit}
  & \varphi_j = \overline{\varphi}_{j} (q_j + q_j^{\dagger}), \\
  & \pi_j = - \frac{i \hbar}{2 \overline{\varphi}_{j}} (q_j - q_j^{\dagger}), \quad \text{with} \nn \\
  & \overline{\varphi}_{j} = \left(\frac{2E_{Cq;j}}{E_{Jq;j}+ 4 E_L}\right)^{1/4}. \nn
\end{align}
Here, $q_j$ ($q_j^{\dagger})$ is the annihilation (creation) operator for qubit $j$ $[j = (1,1), \dots, (N,M)]$ and the effective Josephson energy of the qubits is
$E_{Jq;j}+4E_L$ due to the contributions of the four adjacent inductors.
In terms of these creation and annihilation operators, $H_{0q}$ reads,
\begin{equation}
H_{0q} = \sum_{j=(1,1)}^{(N,M)} \left[ \hbar \omega_j q^{\dagger}_j q_j - \hbar U_j q_j^{\dagger} q_j^{\dagger} q_j q_j \right]
\end{equation}
where we have approximated $\cos(\varphi_{j})$ by its expansion up to fourth order in $\varphi_{j}$ and the transition frequencies $\omega_j$ and nonlinearities $U_j$ are given by
\begin{align}
\omega_j =  & \hbar^{-1} \sqrt{8 E_{Cq;j} (E_{Jq;j} + 4 E_L) }, \quad \text{and}\\
 U_j = &\frac{E_{Cq;j}}{2 \hbar} \frac{E_{Jq;j}}{E_{Jq;j}+ 4 E_L}.
\end{align}
For the SQUID modes we introduce the annihilation (creation) operators $s_{\pm}$ ($s_{\pm}^{\dagger}$) according to,
\begin{align}\label{eq:second-quant-SQUID}
  & \varphi_{\pm;j} = \overline{\varphi}_{\pm} (s_{\pm;j} + s_{\pm;j}^{\dagger}), \\
   & \pi_{\pm;j} =  - \frac{i \hbar}{2 \overline{\varphi}_{\pm}}(s_{\pm;j} - s_{\pm;j}^{\dagger}), \quad \text{with} \nn \\
  &\overline{\varphi}_{\pm} = \left(\frac{\tilde{E}_{C\pm}}{E_L}\right)^{1/4}, \nn
\end{align}
and get 
\begin{equation}
 H_S = \sum_{j=(1,1)}^{(N,M)} \left[ \hbar \omega_+ s_{+;j}^{\dagger} s_{+;j} + \hbar \omega_- s_{-;j}^{\dagger} s_{-;j} \right] ,
\end{equation}
where
\begin{equation} \label{eq:squid-frequs}
\omega_{\pm} = 4 \hbar^{-1}\sqrt{\tilde{E}_{C\pm} E_L}
\end{equation}
and we have neglected the last term of Eq.~(\ref{eq:ham-squids-1}) since $\cos\left(\varphi_{ext}/2\right) \approx - (\varphi_{ac}/2) F(t)$ and the frequencies contained in $F(t)$ strongly differ from $\omega_{+}$ and $\omega_{-}$.
Note that $C_{J} \gg C_{g}$ and thus $\tilde{E}_{C+} \gg \tilde{E}_{C-}$, so that 
$\omega_{+} \gg \omega_{-}$.

\subsection{Accuracy of truncating the transformation}
After having introduced the amplitudes $\overline{\varphi}_{j}$ and $\overline{\varphi}_{\pm}$ in Eqs.~(\ref{eq:second-quant-qubit}) and (\ref{eq:second-quant-SQUID}), we can now estimate the amplitude of the Schrieffer-Wolff generator $S$, see Eqs.~(\ref{eq:SW-main}) or (\ref{eq:SW-app}), to check whether a truncation of the expansion (\ref{eq:SchriefferWolffexpansion}) at fourth order provides a good approximation. For the scenario with at most a single excitation in each qubit and vanishingly small excitation numbers in the SQUID modes, which is of interest here, the generator $S$ has a magnitude given by
\begin{equation} \label{eq:SchriefferWolffamplitude}
\frac{1}{4} \frac{\overline{\varphi}_{j}}{\overline{\varphi}_{\pm}}
\end{equation}
for each site. Typical parameters of our scheme lead to $\overline{\varphi}_{j} \lesssim 0.7$ and $\overline{\varphi}_{\pm} \gtrsim 1.65$ so that $\overline{\varphi}_{j}/4 \overline{\varphi}_{\pm} \approx 0.1$ and the truncation of the expansion in Eq.~(\ref{eq:SchriefferWolffexpansion}) is well justified.

\section{Effective Hamiltonian in the rotating frame} \label{sec:rot-frame}
To analyze which terms of the transformed Hamiltonian (\ref{eq:transformed-ham}) respectively (\ref{eq:transformed-ham-app}) provide the dominant contributions, we now move to a rotating frame, where all qubits rotate at their transition frequencies $\omega_{n,m}$ and the SQUID modes rotate at their oscillation frequencies $\omega_{\pm}$. In transforming to this rotating frame, the raising and lowering operators transform as,
\begin{equation}
q_j \rightarrow e^{-i \omega_j t} q_{j}, \quad \text{and} \quad s_{\pm;j} \rightarrow e^{i \omega_{\pm} t} s_{\pm;j}
\end{equation}
for $j=(1,1), \dots (N,M)$.

\subsection{Adiabatic elimination of the SQUID modes}
For the parameters of our architecture, the SQUID frequencies $\omega_{\pm}$ differ strongly from the qubit transition frequencies $\omega_j$ and the frequencies contained in the driving fields. The SQUID modes $\varphi_{+;n,m}$ and $\varphi_{-;n,m}$ therefore remain to a very good approximation in their ground states, $\ket{0_{+;n,m}}$ and $\ket{0_{-;n,m}}$. They can thus be adiabatically eliminated from the description, where the dominant contributions come from the $\varphi_{-;n,m}$ modes because $\omega_{+} \gg \omega_{-}$. 

The leading order of the adiabatic elimination is obtained by projecting $H_{S}$ and $H_{qS}$ onto the ground states of the SQUIDs, $|0_{\text{SQUIDs}}\rangle= \prod_{n,m=1}^{N,M} |0_{-;n,m},0_{+;n,m}\rangle$. Whereas $H_{S}$ only gives rise to an irrelevant constant, we get for $H_{qS}^{(0)} = \langle 0_{\text{SQUIDs}} | H_{qS} |0_{\text{SQUIDs}} \rangle$ the expression given in Eq.~(\ref{eq:projectGS1}), which we here repeat for completeness
\begin{equation} \label{eq:effective_time_dependent}
\begin{split}
H_{qS}^{(0)} = & -\frac{1}{8} \tilde{E}_J F(t) \sum_{n,m=1}^{N,M} \varphi_{q-;n,m}^2 \\
               & + \frac{1}{16}\frac{1}{4!} \tilde{E}_J F(t) \sum_{n,m=1}^{N,M} \varphi_{q-;n,m}^4 ,
\end{split}
\end{equation}
where,
\begin{align} 
\tilde{E}_{J} & = \chi \, \varphi_{ac} \, E_{J} \quad \text{with} \label{eq:substitute}\\
\chi & = \exp \left(- \frac{\overline{\varphi}_{-;n,m}^2}{2} \right) \label{eq:chidef}
\end{align}
In deriving Eq.~(\ref{eq:effective_time_dependent}) we have used the ground state expectation values
\begin{equation} \label{eq:matrix_elements0}
\begin{split}
 \bra{0_{-;n,m}} \pi_{-;n,m} \ket{0_{-;n,m}} & = 0, \\
 \bra{0_{-;n,m}} \cos (\varphi_{-;n,m}) \ket{0_{-;n,m}} & = \chi, \\
 \bra{0_{-;n,m}} \sin (\varphi_{-;n,m}) \ket{0_{-;n,m}} & = 0. \\
\end{split}
\end{equation}
The Hamiltonian $H_{qS}^{(0)}$ is a sum of contributions from each cell formed by four qubits at its corners which interact via one coupling SQUID.

When considering the terms of one cell only, as described in Sec.~\ref{sec:elim-squid}, and labelling its qubits by indices 1,2,3 and 4, i.e. $(n,m) \to 1, (n+1,m) \to 2, (n+1,m+1) \to 3, (n,m+1) \to 4$,
one can divide the Hamiltonian $H_{qS}^{(0)}$ for one cell into three parts as described in Eq.~(\ref{eq:split-Hamiltonian}),
\begin{equation} \label{eq:split-Hamiltonian-2}
\left. H_{qS}^{(0)} \right|_{\text{cell}}= H_{\text{cell}} =  H_1 + H_2 + H_3,
\end{equation}
where
\begin{equation*}
\begin{split}
 H_1 & = \frac{1}{16} \tilde{E}_{J} F(t) \varphi_1 \varphi_2 \varphi_3 \varphi_4, \\
 H_2 & = - \frac{1}{8} \tilde{E}_{J} F(t) (\varphi_1 + \varphi_4 - \varphi_2 - \varphi_3)^2, \\
 H_3 & = -\tilde{E}_{J} F(t) \sum_{(ijkl)} C_{ijkl} \varphi_i \varphi_j \varphi_k \varphi_l
 \end{split}
\end{equation*}
Here, $H_1$ contains the four-body interactions which will give rise to stabilizer interactions, $H_2$ contains all two-body interactions and $H_3$ all fourth-order interactions beyond the four-body interaction, i.e. terms where at least two of the indices $i,j,k$ and $l$ are the same as we indicate by the notation $\sum_{(ijkl)}$. The coefficients $C_{ijkl}$ have the values $C_{ijkl} = \pm 1/32$ for terms of the from $\varphi_j^{2} \varphi_k \varphi_l$ with $j\neq k$, $j\neq l$ and $k\neq l$,  $C_{ijkl} = \pm 1/64$ for terms of the from $\varphi_j^{2} \varphi_k^{2}$ with $j\neq k$, $C_{ijkl} = \pm 1/96$ for terms of the from $\varphi_j^{3} \varphi_k$ with $j\neq k$ and $C_{ijkl} = \pm 1/384$ for terms of the from $\varphi_j^{4}$. Here, ``$+$'' signs apply whenever an even number of the indices $i,j,k,l$ refer to the same side of the SQUID (i.e. the same diagonal) and ``$-$'' signs otherwise. Whereas we discuss perturbations originating from $H_2$ and $H_3$ in appendix \ref{sec:perturbations}, we now turn to discuss $H_1$ in more detail.

\subsection{Four-body interactions} 
Due to the nonlinearities of the qubits, we can here restrict the description to the two lowest energy levels of each node by replacing $q_j \to \sigma_j^{-}$ ($q_j^{\dag} \to \sigma_j^{+}$). 
The Hamiltonian $H_1$ contains 16 terms of the form $S_1 S_2 S_3 S_4$, where $S_j \in \{ e^{-i \omega_j t} \sigma^-_j, e^{i \omega_j t} \sigma_j^{+} \}$ (see \ref{eq:split-Hamiltonian-2}). Ignoring the oscillating pre-factor $F(t)$, these terms rotate at the angular frequencies
as listed in table \ref{tab:1}.
\begin{table}[ht]
\centering
\begin{tabular}{|l|c||l|c|}
\hline
term
& frequency 
& term
& frequency
\\
\hline
$\sigma_1^+ \sigma_2^+ \sigma_3^+ \sigma_4^+$ & $\nu_{1}$ &   $\sigma_1^+ \sigma_2^+ \sigma_3^+ \sigma_4^-$ & $\nu_{5}$\\
$\sigma_1^- \sigma_2^- \sigma_3^- \sigma_4^-$ & $-\nu_{1}$ &  $\sigma_1^+ \sigma_2^+ \sigma_3^- \sigma_4^+$ & $\nu_{6}$\\
$\sigma_1^+ \sigma_2^+ \sigma_3^- \sigma_4^-$ & $\nu_{2}$ &  $\sigma_1^+ \sigma_2^- \sigma_3^+ \sigma_4^+$ & $\nu_{7}$\\
$\sigma_1^+ \sigma_2^- \sigma_3^+ \sigma_4^-$ & $\nu_{3}$ &  $\sigma_1^- \sigma_2^+ \sigma_3^+ \sigma_4^+$ & $\nu_{8}$\\
$\sigma_1^+ \sigma_2^- \sigma_3^- \sigma_4^+$ & $\nu_{4}$ &  $\sigma_1^- \sigma_2^- \sigma_3^- \sigma_4^+$ & $-\nu_{5}$\\
$\sigma_1^- \sigma_2^+ \sigma_3^+ \sigma_4^-$ & $-\nu_{4}$ &  $\sigma_1^- \sigma_2^- \sigma_3^+ \sigma_4^-$ & $-\nu_{6}$\\
$\sigma_1^- \sigma_2^+ \sigma_3^- \sigma_4^+$ & $-\nu_{3}$ &  $\sigma_1^- \sigma_2^+ \sigma_3^- \sigma_4^-$ & $-\nu_{7}$\\
$\sigma_1^- \sigma_2^- \sigma_3^+ \sigma_4^+$ & $-\nu_{2}$ &  $\sigma_1^+ \sigma_2^- \sigma_3^- \sigma_4^-$ & $-\nu_{8}$\\
\hline
\multicolumn{4}{l}{$\,$}\\
\multicolumn{4}{l}{frequencies:}\\
\hline
\multicolumn{4}{l}{$\nu_{1} = \omega_1 + \omega_2 + \omega_3 + \omega_4$}\\
\multicolumn{4}{l}{$\nu_{2} = \omega_1 + \omega_2 - \omega_3 - \omega_4$}\\
\multicolumn{4}{l}{$\nu_{3} = \omega_1 - \omega_2 + \omega_3 - \omega_4$}\\
\multicolumn{4}{l}{$\nu_{4} = \omega_1 - \omega_2 - \omega_3 + \omega_4$}\\
\multicolumn{4}{l}{$\nu_{5} = \omega_1 + \omega_2 + \omega_3 - \omega_4$}\\
\multicolumn{4}{l}{$\nu_{6} = \omega_1 + \omega_2 - \omega_3 + \omega_4$}\\
\multicolumn{4}{l}{$\nu_{7} = \omega_1 - \omega_2 + \omega_3 + \omega_4$}\\
\multicolumn{4}{l}{$\nu_{8} = -\omega_1 + \omega_2 + \omega_3 + \omega_4$}\\  
\end{tabular}
\caption{\label{tab:1} The four-body terms contained in $H_1$ of Eq.~(\ref{eq:split-Hamiltonian-2}) and their rotation frequencies.}
\end{table}
Denoting the sum of the terms in the first column of table \ref{tab:1} by $V_{1}$ and the sum of the terms in the second column by $V_2$ it can readily be verified that $V_1+V_2=A_S=\sigma^x_1\sigma^x_2\sigma^x_3\sigma^x_4$ and $V_1-V_2=B_P=\sigma^y_1\sigma^y_2\sigma^y_3\sigma^{y}_{4}$. Motivated by the specific form of the Hamiltonian $H_1$, see Eq.~(\ref{eq:split-Hamiltonian-2}), we thus define two oscillating fields $F_{X}$ and $F_{Y}$,
\begin{equation} \label{eq:fields}
 \begin{split}
  & F_X(t) = f_{1,1,1,1} + f_{1,1,-1,-1} + f_{1,-1,1,-1} + f_{1,-1,-1,1} \\ 
  & F_Y(t) = f_{1,1,1,-1} + f_{1,1,-1,1} + f_{1,-1,1,1} + f_{-1,1,1,1} 
 \end{split}
\end{equation}
with $f_{a,b,c,d} = \cos[(a \omega_1 + b \omega_2 + c \omega_3 + d \omega_4)t]$. Each of the terms in $F_X$ ($F_{Y}$) brings two of the four-body processes in the first (second) column of table \ref{tab:1} into resonance (makes them non-rotating). Next to resonant terms, each field also generates rotating terms. The rotation frequency of these terms is the detuning between different components of $F_{X}$ and $F_{Y}$. As the least detuning is in the GHz range ($\sim$ 0.4 GHz) and the strength of four-body interaction is in the MHz range ($\sim$ 1 MHz), the rotating terms can be neglected in a RWA.  Hence the external oscillating field $F_s = F_X + F_Y$ gives rise to the star interaction $A_s$ and the external field $F_p = F_X - F_Y$ gives rise to the plaquette interaction $B_p$, 
\begin{equation} \label{eq:fields2}
F(t) = \begin{cases} 
      F_s = F_X + F_Y \, \Rightarrow \, H_1 \propto A_s = \sigma_1^x \sigma_2^x \sigma_3^x \sigma_4^x \\
      F_p = F_X - F_Y  \, \Rightarrow \, H_1 \propto  B_p = \sigma_1^y \sigma_2^y \sigma_3^y \sigma_4^y
   \end{cases}
\end{equation}
and the strength of the stabilizer interaction is
\begin{equation}
 J_s = J_p = \frac{1}{16} \tilde{E}_{J} \prod_{i=1}^4 \overline{\varphi}_{i}
\end{equation}
Besides these stabilizer interactions, there are also two-body interactions between the qubits which may lead to corrections in the effective Hamiltonian (\ref{eq:toric-code-ham}). We now turn to estimate their effect.

\section{Perturbations} \label{sec:perturbations}
In this section we estimate the effects of the remaining terms of the transformed Hamiltonian $\tilde{H}$, see Eq.~(\ref{eq:transformed-ham}) or (\ref{eq:transformed-ham-app}). These include contributions contained in the Hamiltonians $H_{Iq}$, see Eq.~(\ref{eq:qubit-int}), and $H_{qS}$, see Eq.~(\ref{eq:nonlinear}). In our approach, the transition frequencies of the qubits $\omega_{n,m}$ and the frequency spectrum of the oscillating fluxes $\nu_{k}$ are chosen such the all undesired terms are highly off-resonant and therefore strongly suppressed. They however lead to a frequency shift
\begin{equation} \label{eq:shifts}
\Delta_{j} = \Delta_{j}^{(1)} + \Delta_{j}^{(2)} +\Delta_{j}^{(3)} +\Delta_{j}^{(4)}
\end{equation}
for each qubit $j$, where $\Delta_{j}^{(1)}, \Delta_{j}^{(2)}, \Delta_{j}^{(3)}$ and $\Delta_{j}^{(4)}$ are given in Eqs.~(\ref{eq:shift1}), (\ref{eq:shift2}),(\ref{eq:shift3}) and (\ref{eq:shift4}). These frequency shifts can easily be compensating for by modifying the frequency spectrum of the applied driving fields accordingly, so that we did not include them in our preceding analysis. They can be calculated via the following time dependent perturbation theory.

In the frame where all qubits rotate at their transition frequencies, the transformed Hamiltonian $\tilde{H}$
can be written as
\begin{equation}\label{eq:H2-frequ-decomp}
\tilde{H} = H_{\text{TC}} + \sum_{\alpha} \hat{h}_{\alpha} e^{i\omega_{\alpha}t},
\end{equation}
where $H_{\text{TC}}$ is the Toric Code Hamiltonian as introduced in Eq.~(\ref{eq:toric-code-ham}) and the remaining terms have been written as a sum of all their frequency components. Hence $\hat{h}_{\alpha}$ is the sum of all terms that rotate as $e^{i\omega_{\alpha}t}$, where $\omega_{\alpha}$ contains all contributing frequencies, i.e. the oscillation frequencies of the operators and, for terms with a prefactor $F(t)$, the frequency of the oscillating flux bias. Note that both, $\omega_{\alpha} > 0$ and $\omega_{\alpha} < 0$ contribute to the sum in Eq.~(\ref{eq:H2-frequ-decomp}) since $\tilde{H}$ is Hermitian.

Following approaches outlined in \cite{James00,Goldman14,solid_NMR}, the dynamics generated by a Hamiltonian of the form as in Eq.~(\ref{eq:H2-frequ-decomp}) with $||\hat{h}_{\alpha}|| \ll \hbar \omega_{\alpha}$ can be approximated by a time independent effective Hamiltonian that is an expansion in powers of the interaction strength over oscillation frequency \cite{solid_NMR},
\begin{equation} \label{eq:effham-expand}
H_{\text{eff}} = H_{\text{eff},0} + H_{\text{eff},1} + H_{\text{eff},2} + H_{\text{eff},3} + \dots ,
\end{equation} 
where the individual contributions read
\begin{align*}
H_{\text{eff},0} & = H_{\text{TC}} \\
H_{\text{eff},1} & = \sum_{\alpha,\beta} \frac{\delta(\omega_{\alpha}+\omega_{\beta})}{2 \hbar \omega_{\beta}} \left[\hat{h}_{\alpha},\hat{h}_{\beta} \right] \\
H_{\text{eff},2} & = \sum_{\alpha,\beta} \frac{\delta(\omega_{\alpha}+\omega_{\beta})}{2 \hbar^2 \omega_{\alpha} \omega_{\beta}} \left[\left[H_{\text{TC}},\hat{h}_{\alpha} \right] ,\hat{h}_{\beta} \right]\\
& + \sum_{\alpha,\beta,\gamma} \frac{\delta(\omega_{\alpha}+\omega_{\beta}+\omega_{\gamma})}{3 \hbar^2 \omega_{\beta} \omega_{\gamma}} \left[\left[\hat{h}_{\alpha},\hat{h}_{\beta} \right] ,\hat{h}_{\gamma} \right]\\
H_{\text{eff},3} & = \sum_{\alpha,\beta,\gamma} \frac{\delta(\omega_{\alpha}+\omega_{\beta}+\omega_{\gamma})}{3 \hbar^3 \omega_{\alpha} \omega_{\beta} \omega_{\gamma}} \left[\left[\left[H_{\text{TC}},\hat{h}_{\alpha} \right] ,\hat{h}_{\beta} \right],\hat{h}_{\gamma} \right]\\
& + \sum_{\alpha,\beta,\gamma,\delta} \frac{\delta(\omega_{\alpha}+\omega_{\beta}+\omega_{\gamma}+\omega_{\delta})}{4 \hbar^3 \omega_{\beta} \omega_{\gamma} \omega_{\delta}} \left[\left[\left[\hat{h}_{\alpha},\hat{h}_{\beta} \right] ,\hat{h}_{\gamma} \right] ,\hat{h}_{\delta} \right]
\end{align*} 
and $\delta(\omega)$ is the Dirac $\delta$-function.

Since $||\hat{h}_{\alpha}|| \ll \hbar \omega_{\alpha}$, the first term of $H_{\text{eff},2}$ and the first term of $H_{\text{eff},3}$ are at least two orders of magnitude smaller than $H_{\text{TC}}$ and we neglect them. We now turn to discuss the other perturbations to $H_{\text{TC}}$ one by one.

\subsection{Corrections to the adiabatic elimination of the SQUIDs}
\label{sec:first-order-corrs-adiabat}
Besides the terms contained in $H_{qs}^{(0)}$ that are found by projecting the Hamiltonian $\tilde{H}$ onto the ground states of the SQUID modes, there are first-order corrections to the adiabatic elimination. These are processes mediated by the creation and annihilation of a virtual excitation in a SQUID mode.
To estimate their effect, we first note that terms proportional to $\cos (\varphi_{-;n,m})$ do not create single excitations in the SQUID modes since $ \bra{0_{-;n,m}} \cos (\varphi_{-;n,m}) \ket{1_{-;n,m}} = 0$ and calculate the matrix elements
\begin{equation} \label{eq:matrix_elements1}
\begin{split}
 \bra{0_{-;n,m}} \sin (\varphi_{-;n,m}) \ket{0_{-;n,m}} & = 0, \\
 \bra{0_{-;n,m}} \sin (\varphi_{-;n,m}) \ket{1_{-;n,m}} & = \chi,\\
  \bra{1_{-;n,m}} \sin (\varphi_{-;n,m}) \ket{1_{-;n,m}} & = 0.
\end{split}
\end{equation}
In the subspace of at most one excitation, we may thus replace
\begin{equation} \label{eq:adiabat-correct}
\sin (\varphi_{-;n,m}) \to \chi (s_{-;n,m} + s_{-;n,m}^\dag ) = \frac{\chi}{\overline{\varphi}_-} \varphi_{-;n,m}, 
\end{equation}
to approximate $\Delta H_{qS}= H_{qS}-H_{qS}^{(0)}$, c.f. Eq.~(\ref{eq:split-Hamiltonian-2}), as
\begin{align}
  \Delta H_{qS}  \approx  &-4 \sum_{n,m=1}^{N,M} \frac{E_{Cq;n,m}}{\hbar^{2}} \pi_{n,m} \pi_{S-;n,m}\\
             &- \frac{\tilde{E}_J}{2 \overline{\varphi}_-} F(t) \sum_{n,m=1}^{N,M} \varphi_{-;n,m} \varphi_{q-;n,m}\nn 
\end{align}
where we have neglected the terms in Eq.~(\ref{eq:nonlinear}) that are proportional to $\varphi_{q-;n,m}^{3}$ as these are a factor $\overline{\varphi}_{n,m}^{2}/24 \ll 1$ smaller than the terms that are linear in $\varphi_{q-;n,m}$. 
We now consider corrections to $H_{TC}$ coming from second order processes of $\Delta H_{qS}$.

Using the formulation in second quantization, see Eqs.~(\ref{eq:second-quant-qubit}) and (\ref{eq:second-quant-SQUID}), one finds for the first term of $\Delta H_{qS}$
that the perturbation theory of Eq.~(\ref{eq:effham-expand}) is applicable if
\begin{equation}\label{eq:rotwave-cond-1}
\left| \frac{E_{Cq;j}}{\hbar (\omega_{-} - \omega_{j}) \overline{\varphi}_{j} \overline{\varphi}_{-}} \right| \ll 1.
\end{equation}
For this regime, the effective Hamiltonian $H_{\text{eff},1}$ according to Eq.~(\ref{eq:effham-expand}) for the first term of $\Delta H_{qS}$ just contains a frequency shift
\begin{equation}\label{eq:shift1}
\Delta_{j}^{(1)} = - \frac{4 E_{Cq;j}^{2}}{\hbar^{2} (\omega_{-} - \omega_{j}) \overline{\varphi}_{j}^{2} \overline{\varphi}_{-}^{2}} .
\end{equation}
for each qubit $j = (n,m)$. Here, the prefactor 4 accounts for the fact that each qubit has 4 neighboring SQUIDs. For the set of parameters discussed in Sec.~\ref{sec:parameters}, we find for the qubits with high transition frequencies $E_{Cq;j}/(h \overline{\varphi}_{j} \overline{\varphi}_{-}) \sim 500\,\text{MHz}$ leading to $\Delta_{j}^{(1)}\sim 200\,\text{MHz}$, i.e. a shift of $50\,\text{MHz}$ from each coupling to a SQUID. For qubits with low transition frequencies the shifts are $\Delta_{j}^{(1)}\sim 100\,\text{MHz}$, i.e. a shift of $25\,\text{MHz}$ from each coupling to a SQUID. 
Besides frequency shifts, the first term of $\Delta H_{qS}$ can also lead to interactions between two qubits of one cell. These are ineffective provided any pair of qubits in a cell is sufficiently detuned from each other,
\begin{equation}
\left| \frac{E_{Cq;j}E_{Cq;l}}{\hbar^{2} (\omega_{l} - \omega_{j}) (\omega_{-} - \omega_{j}) \overline{\varphi}_{j} \overline{\varphi}_{l} \overline{\varphi}_{-}^{2}} \right| \ll 1,
\end{equation}
which holds since any pair of qubits in one cell is at least detuned by $700\,\text{MHz}$ from each other, see table \ref{tab:transition-frequs}.
Note that the detuning between two neighboring qubits on a diagonal $n$ that would couple via two SQUIDs is significantly higher ($\sim 10\,\text{GHz}$).
The next higher order non-vanishing term of the expansion (\ref{eq:effham-expand}) is here the second term of $H_{\text{eff},3}$ which is more than two orders of magnitude smaller than $\Delta_{j}^{(1)}$.

In a similar way as second-order terms can lead to qubit-qubit interactions in one cell, fourth-order terms could mediate interactions between two qubits in neighboring cells. For our parameters, these interactions would have a strength of $\sim 0.5\,\text{MHz}$ but are ineffective as qubits in neighboring cells are always detuned by at least $100\,\text{MHz}$.

Via an analogous calculation, one finds for the second term of $\Delta H_{qS}$ that the perturbation theory of Eq.~(\ref{eq:effham-expand}) is applicable if
\begin{equation} \label{eq:rotwave-cond-2}
\left| \frac{\tilde{E}_{J} \overline{\varphi}_{j}}{ 2 \hbar [(\omega_{-} \pm \omega_{j}) \pm \nu_{k}]} \right| \ll 1
\end{equation}
and leads to the frequency shifts (taking into account that each qubit couples to 4 SQUIDs)
\begin{equation}\label{eq:shift2}
\Delta_{j}^{(2)} = -2\frac{\tilde{E}_{J}^{2} \overline{\varphi}_{j}^{2} }{\hbar^{2}} \sum_{k=1}^8 \sum_{x=0,1} \frac{\omega_- + (-1)^{x}\omega_{j}}{[\omega_- + (-1)^{x}\omega_{j}]^{2} - \nu_{k}^{2}}
\end{equation}
for each qubit $j$, where the frequencies $\nu_{k}$ that are contained in the driving field $F(t)$ are defined in table \ref{tab:1}. For the set of parameters as given in Sec.~\ref{sec:parameters}, we find $|\Delta_{j}^{(2)}| \lesssim 20\,\text{MHz}$.
%
%
%
%
%
%
The next higher order non-vanishing term of the expansion (\ref{eq:effham-expand}) is here the second term of $H_{\text{eff},3}$ which is more than two orders of magnitude smaller than $\Delta_{j}^{(2)}$.

\subsection{Perturbations contained in $H_{Iq}$}
The transformed Hamiltonian $\tilde{H}$, see Eq.~(\ref{eq:transformed-ham}) or (\ref{eq:transformed-ham-app})
contains residual interactions between neighboring qubits on each diagonal $n$ as described in Eq.~(\ref{eq:qubit-int}). Their strength is $2 E_{L} \overline{\varphi}_{n,m} \overline{\varphi}_{n,m+1}$ for the interaction between qubits $(n,m)$ and $(n,m+1)$ on diagonal $n$. Hence provided that
\begin{equation} \label{eq:rotwave-cond-3}
\left| \frac{2 E_{L} \overline{\varphi}_{n,m} \overline{\varphi}_{n,m+1}}{\hbar (\omega_{n,m} - \omega_{n,m+1})} \right| \ll  1 
\end{equation}
the perturbation theory of Eq.~(\ref{eq:effham-expand}) is applicable and leads to the frequency shift
\begin{equation}\label{eq:shift3}
\Delta_{n,m}^{(3)} = \frac{4 E_{L}^{2} \overline{\varphi}_{n,m}^{2} \overline{\varphi}_{n,m-1}^{2}}{\hbar^{2} (\omega_{n,m} - \omega_{n,m-1})} + \frac{4 E_{L}^{2} \overline{\varphi}_{n,m}^{2} \overline{\varphi}_{n,m+1}^{2}}{\hbar^{2} (\omega_{n,m} - \omega_{n,m+1})},
\end{equation}
for qubit $(n,m)$. For the parameters discussed in Sec.~\ref{sec:parameters}, we find $2 E_{L} \overline{\varphi}_{n,m} \overline{\varphi}_{n,m+1} / h \approx 300\,\text{MHz}$ whereas neighboring qubits on a diagonal are detuned by almost $10\,\text{GHz}$ so that 
$|\Delta_{n,m}^{(3)}| \sim 18 \,\text{MHz}$. In a higher order of the perturbation theory of Eq.~(\ref{eq:effham-expand}), each qubit $(n,m)$ can also mediate an interaction between its two neighboring qubits on the diagonal $n$ due to the couplings $2 E_{L}\varphi_{n,m-1} \varphi_{n,m}$ and $2 E_{L}\varphi_{n,m} \varphi_{n,m+1}$. To suppress these interactions of strength $10\,\text{MHz}$, next-nearest neighbor qubits need to be sufficiently detuned from one another,
\begin{equation} \label{eq:rotwave-cond-3a}
\left| \frac{4 E_{L}^{2} \overline{\varphi}_{n,m-1} \overline{\varphi}_{n,m}^{2} \overline{\varphi}_{n,m+1}}{\hbar^{2} (\omega_{n,m} - \omega_{n,m+1}) (\omega_{n,m-1} - \omega_{n,m+1})} \right| \ll  1 ,
\end{equation}
which requires $|\omega_{n,m-1} - \omega_{n,m+1}| \gtrsim 2 \pi \times 100\,\text{MHz}$ as fulfilled by the frequencies in table \ref{tab:transition-frequs}. 

Furthermore, there could be interactions between qubits in neighboring cells, say qubit $(n,m)$ and qubit $(n,m+2)$, since qubits $(n,m)$ and qubit $(n,m+1)$ couple via $2 E_{L}\varphi_{n,m} \varphi_{n,m+1}$ and qubits $(n,m+1)$ and qubit $(n,m+2)$ couple in a second order process via their interaction with a common SQUID as discussed in Sec.~\ref{sec:first-order-corrs-adiabat}. This third-order interactions are strongly suppressed since their strength is $\sim 3\,\text{MHz}$ and thus 30 times smaller than the detuning between the respective qubits.

\subsection{Residual two-body interactions in $H_{qS}^{(0)}$}
\label{sec:residual-2body-app}
When projected onto the ground states of the SQUID modes in the zeroth order of adiabatic elimination, the Hamiltonian  $H_{qS}^{(0)} = \langle 0_{\text{SQUIDs}} | H_{qS} |0_{\text{SQUIDs}} \rangle$ contains terms that are quadratic in $\varphi_{q-;n,m}$ and thus contain two-body interactions, see Eq.~(\ref{eq:effective_time_dependent}).
In the rotating frame, the two-body interaction Hamiltonian of one cell as given by $H_2$ in Eq.~(\ref{eq:split-Hamiltonian-2}) can be written as,
\begin{equation} \label{eq:two_body_rotating}
\begin{split}
    H_2  = & \sum_{j,l=1}^4 \sum_{k=1}^8 B_{j,l} \left\{ \left[e^{i (\omega_j - \omega_l + \nu_{k})t}+e^{i (\omega_j - \omega_l - \nu_{k})t}\right] q_j^{\dagger} q_l \right.\\
      & \quad + \left. \left[ e^{i (\omega_j + \omega_l + \nu_{k})t}+e^{i  (\omega_j + \omega_l - \nu_{k})t} \right] q_j^{\dagger} q_l^{\dagger} + \text{H.c.}\right\},
\end{split}
\end{equation}
where the coupling coefficients read $B_{j,l} = \mp \tilde{E}_{J}\overline{\varphi}_{j}\overline{\varphi}_{l}/8$, where the ``$-$'' sign applies for $j=l$ and $(j,l) =  (1,4)$ or $(2,3)$, and the ``$+$'' sign applies otherwise. The perturbation theory of Eq.~(\ref{eq:effham-expand}) applies to $H_{2}$ provided that
\begin{equation} \label{eq:rotwave-cond-4}
\left| \frac{\tilde{E}_{J}\overline{\varphi}_{j}\overline{\varphi}_{l}}{8 \hbar (\omega_j \pm \omega_l \pm \nu_{k})}\right| \ll 1
\end{equation}
for all $j,l$ and $k$, c.f. Eq.~(\ref{eq:two_body_rotating}).
Hence the spectrum of the driving fields must be chosen such that conditions (\ref{eq:rotwave-cond-2}) and (\ref{eq:rotwave-cond-4}) are met.

To estimate the corrections due to $H_{2}$ as written in Eq.~(\ref{eq:two_body_rotating}), one thus needs to consider the commutation relations,
\begin{align} \label{eq:comm_rel}
[q_j q_l^{\dagger}, q_j^{\dagger} q_l] & = (1-\delta_{j,l}) \left(q_l^{\dagger} q_l - q_j^{\dagger} q_j\right), \\
[q_j q_l, q_j^{\dagger} q_l^{\dagger}] & = (1+\delta_{j,l}) \left(q_j^{\dagger} q_{j} + q_l^{\dagger} q_l + 1 \right), \nn \\
[q_j q_l, q_j q_l^{\dagger}] & = (1+\delta_{j,l}) \, q_j q_j, \nn \\
[q_j q_l^{\dagger}, q_j^{\dagger} q_l^{\dagger} ] & = (1+\delta_{j,l}) \, q_l^{\dagger} q_l^{\dagger}, \nn 
\end{align} 
as these are the only ones that, together with their oscillating prefactors $F_{s}$ or $F_{p}$, lead to non-rotating terms. The two commutators in the last two lines of Eq.~(\ref{eq:comm_rel}) would lead to transitions out of the qubit subspace and can thus be ignored because of the nonlinearity of the qubits.

Hence, to leading order the Hamiltonian $H_{2}$ will lead to shifts of the transition frequencies of the qubits. Taking into account that each cell is embedded into the total lattice and hence each qubit experiences frequency shifts from the two-body interactions in all four cells it is part of, we find that the qubit transition frequencies are shifted by
\begin{align} \label{eq:shift4}
\Delta_{j}^{(4)} & = - \frac{\tilde{E}_{J}^{2}}{8}  \sum_{\{l:|j-l|=1\}}\sum_{k=1}^8 \overline{\varphi}_{j}^{4}\frac{2 \omega_{j}}{4 \omega_{j}^{2} - \nu_{k;j,l}^{2}} \\
& -\frac{\tilde{E}_{J}^{2}}{32}  \sum_{\{l:|j-l|=1\}}\sum_{k=1}^8 \frac{ \overline{\varphi}_{j}^{2}\overline{\varphi}_{l}^{2} [\omega_{j} + \omega_{l}]}{[\omega_{j} + \omega_{l}]^{2} - \nu_{k;j,l}^{2}}, \nn \\
& +\frac{\tilde{E}_{J}^{2}}{32}  \sum_{\{l:|j-l|=1\}}\sum_{k=1}^8 \frac{ \overline{\varphi}_{j}^{2}\overline{\varphi}_{l}^{2} [\omega_{j} - \omega_{l}]}{[\omega_{j} -\omega_{l}]^{2} - \nu_{k;j,l}^{2}}, \nn
\end{align}
where the sum $\sum_{\{l:|j-l|=1\}}$ runs over all qubits $l$ that are nearest neighbors to qubit $j$ and $\nu_{k;j,l}$ denotes the drive frequencies at the SQUID that connects qubits $j$ and $l$. For the set of parameters as given in Sec.~\ref{sec:parameters}, we find $|\Delta_{j}^{(4)}| \lesssim 30\,\text{MHz}$.
%
%
%
%
%

\section{Expected Measurement Outcomes}\label{sec:outcomes}
In this section we discuss a set of possible measurement out comes for the topologically ordered states on a eight-qubit lattice, c.f. Fig.~\ref{fig:min-lattice}.
One ground state of the Toric Code Hamiltonian on this lattice can be written as 
\begin{widetext}
\begin{equation}
\ket{\psi_{1}}=  \frac{1}{4\sqrt{2}} (\mathbbm{1}+ \sigma_{1}^y \sigma_{4}^y \sigma_{5}^y \sigma_{3}^y) (\mathbbm{1}+ \sigma_{2}^y \sigma_{3}^y \sigma_{6}^y \sigma_{4}^y)(\mathbbm{1}+ \sigma_{5}^y \sigma_{8}^y\sigma_{1}^y \sigma_{7}^y)(\mathbbm{1}+ \sigma_{6}^y \sigma_{7}^y \sigma_{2}^y \sigma_{8}^y)\prod_{j=1}^8 \ket{+_j},
\end{equation}
\end{widetext}
where $\ket{+_{j}}$ is the eigenstate of $\sigma_{j}^x$ with eigenvalue 1, $\sigma_{j}^x\ket{+_{j}} = \ket{+_{j}}$.
The other three ground states can be found by applying the loop operators $\sigma_{5}^{y}\sigma_{6}^{y}$, $\sigma_{4}^{y}\sigma_{8}^{y}$ and $\sigma_{4}^{y}\sigma_{8}^{y} \sigma_{5}^{y}\sigma_{6}^{y}$,
\begin{align}
\ket{\psi_{2}} & = \sigma_{5}^{y}\sigma_{6}^{y}\ket{\psi_{1}}\\
\ket{\psi_{3}} & = \sigma_{4}^{y}\sigma_{8}^{y}\ket{\psi_{1}}\nn \\
\ket{\psi_{4}} & = \sigma_{4}^{y}\sigma_{8}^{y} \sigma_{5}^{y}\sigma_{6}^{y}\ket{\psi_{1}}. \nn
\end{align} 
Whereas the reduced density matrices of individual spins are all completely mixed (i.e. proportional to an identity matrix) for all these four states, they can be distinguished via their two spin correlations, see table \ref{tab:two-spi-corrs}.
\begin{table*}[t!]
\noindent
\centering
\begin{tabular}{|c||c|c|c|c|c|c|c|c|}
\hline
& $\langle \sigma_{1}^{x} \sigma_{2}^{x} \rangle$
& $\langle \sigma_{3}^{x} \sigma_{4}^{x} \rangle$
& $\langle \sigma_{5}^{x} \sigma_{6}^{x} \rangle$
& $\langle \sigma_{7}^{x} \sigma_{8}^{x} \rangle$
& $\langle \sigma_{3}^{x} \sigma_{7}^{x} \rangle$
& $\langle \sigma_{1}^{x} \sigma_{5}^{x} \rangle$
& $\langle \sigma_{4}^{x} \sigma_{8}^{x} \rangle$
& $\langle \sigma_{2}^{x} \sigma_{6}^{x} \rangle$
\\
\hline
$\ket{\psi_{1}}$ & 0 & 1 & 0 & 1 & 0 & 1 & 0 & 1 \\
$\ket{\psi_{2}}$ & 0 & 1 & 0 & 1 & 0 & -1 & 0 & -1 \\
$\ket{\psi_{3}}$ & 0 & -1 & 0 & -1 & 0 & 1 & 0 & 1 \\
$\ket{\psi_{4}}$ & 0 & -1 & 0 & -1 & 0 & -1 & 0 & -1 \\
\hline
\end{tabular}
\caption{\label{tab:two-spi-corrs} Two-qubit correlations along non-contractible loops for the 4 ground states of the Toric Code on an 8-qubit lattice. }
\end{table*}



\begin{thebibliography}{55}


\bibitem{Stormer99}
H. L. Stormer, D. C. Tsui, and A. C. Gossard,
{\it The fractional quantum Hall effect},
Rev. Mod. Phys. {\bf 71}, 298 (1999).

\bibitem{Balents10}
L. Balents,
{\it Spin liquids in frustrated magnets},
Nature {\bf 464}, 199 (2010).

\bibitem{Nayak08}
C. Nayak, S. H. Simon, A. Stern, M. Freedman, and S. Das Sarma,
{\it Non-Abelian anyons and topological quantum computation},
Rev. Mod. Phys. {\bf 80}, 1083 (2008).

\bibitem{Kitaev2003}
A. Kitaev, 
{\it Fault-tolerant quantum computation by anyons},
Ann. Phys. {\bf 303}, 2 (2003).

\bibitem{Kitaev09}
A. Kitaev and C. Laumann,
{\it Topological phases and quantum computation},
arXiv:0904.2771

\bibitem{Fowler12}
A. G. Fowler, M. Mariantoni, J. M. Martinis, and A. N. Cleland,
{\it Surface codes: Towards practical large-scale quantum computation},
Phys. Rev. A {\bf 86}, 032324 (2012).

\bibitem{Dennis02}
E. Dennis, A. Kitaev, A. Landahl, and J. Preskill,
{\it Topological quantum memory}
J. Math. Phys. {\bf 43}, 4452 (2002).

\bibitem{Bombin13}
H. Bombin, R. W. Chhajlany, M. Horodecki and M. A. Martin-Delgado,
{\it Self-correcting quantum computers},
New J. Phys. {\bf 15}, 055023 (2013).

\bibitem{convention}
The convention we use here is motivated by the usual notation for superconducting qubits but differs from conventions mostly used when discussing computational aspects of the Toric Code.

\bibitem{Ioffe02}
L. B. Ioffe, M. V. Feigel'man, A. Ioselevich, D. Ivanov, M. Troyer, and G. Blatter,
{\it Topologically protected quantum bits using Josephson junction arrays,}
Nature {\bf 415}, 503 (2002)

\bibitem{Gladchenko09}
S. Gladchenko, D. Olaya, E. Dupont-Ferrier, B. Doucot, L. B. Ioffe, M. E. Gershenson,
{\it Superconducting nanocircuits for topologically protected qubits},
Nat. Phys. {\bf 5}, 48 (2009).
 
\bibitem{Yao12}
X.-C. Yao, T.-X. Wang, H.-Z. Chen, W.-B. Gao, A. G. Fowler, R. Raussendorf, Z.-B. Chen, N.-L. Liu, C.-Y. Lu, Y.-J. Deng, Y.-A. Chen, and J.-W. Pan,
{\it Experimental demonstration of topological error correction,}
Nature {\bf 482}, 489 (2012).

\bibitem{Nigg14}
D. Nigg, M. M\"uller, E. A. Martinez, P. Schindler, M. Hennrich, T. Monz, M. A. Martin-Delgado, R. Blatt,
{\it Quantum computations on a topologically encoded qubit,}
Science {\bf 345}, 302 (2014). 

\bibitem{Kitaev06}
A. Kitaev,
{\it Anyons in an exactly solved model and beyond},
Ann. Phys. {\bf 321}, 2 (2006).

\bibitem{Chun15}
S. H. Chun, J.-W. Kim, J. Kim, H. Zheng, C. C. Stoumpos, C. D. Malliakas, J. F. Mitchell, K. Mehlawat, Y. Singh, Y. Choi, T. Gog, A. Al-Zein, M. Moretti Sala, M. Krisch, J. Chaloupka, G. Jackeli, G. Khaliullin, and B. J. Kim,
{\it Direct evidence for dominant bond-directional interactions in a honeycomb lattice iridate Na2IrO3},
Nat. Phys. {\bf 11}, 462 (2015).

\bibitem{Banerjee16}
A. Banerjee, C. A. Bridges,	J.-Q. Yan,	A. A. Aczel, L. Li,	M. B. Stone, G. E. Granroth, M. D. Lumsden,	Y. Yiu,	J. Knolle, S. Bhattacharjee, D. L. Kovrizhin, R. Moessner,	D. A. Tennant, D. G. Mandrus, and S. E. Nagler,
{\it Proximate Kitaev quantum spin liquid behaviour in a honeycomb magnet},
Nat. Mater. {\bf 15}, 733 (2016).

\bibitem{Kelly14}
J. Kelly, R. Barends, A. G. Fowler, A. Megrant, E. Jeffrey, T. C. White, D. Sank, J. Y. Mutus, B. Campbell, Yu Chen, Z. Chen, B. Chiaro, A. Dunsworth, I.-C. Hoi, C. Neill, P. J. J. O'Malley, C. Quintana, P. Roushan, A. Vainsencher, J. Wenner, A. N. Cleland, and J. M. Martinis,
{\it State preservation by repetitive error detection in a superconducting quantum circuit},
Nature {\bf 519}, 66 (2015).
%
\bibitem{Riste14}
D. Rist\`e, S. Poletto, M.-Z. Huang, A. Bruno, V. Vesterinen, O.-P. Saira, and L. DiCarlo,
{\it Detecting bit-flip errors in a logical qubit using stabilizer measurements},
Nat. Commun. {\bf 6}, 6983 (2015).

\bibitem{Corcoles15}
A. D. C\'{o}rcoles, Easwar Magesan, Srikanth J. Srinivasan, Andrew W. Cross, M. Steffen, Jay M. Gambetta, and Jerry M. Chow,
{\it Detecting arbitrary quantum errors via stabilizer measurements on a sublattice of the surface code},
Nat. Comm. {\bf 6}, 6979 (2015).

\bibitem{Blumoff16}
J. Z. Blumoff, K. Chou, C. Shen, M. Reagor, C. Axline, R.T. Brierley, M. P. Silveri, C. Wang, B. Vlastakis, S. E. Nigg, L. Frunzio, M. H. Devoret, L. Jiang, S. M. Girvin, and R. J. Schoelkopf,
{\it Implementing and characterizing precise multi-qubit measurements},
Phys. Rev. X {\bf 6}, 031041  (2016).

\bibitem{Marcos13}
D. Marcos, P. Rabl, E. Rico, and P. Zoller,
{\it Superconducting Circuits for Quantum Simulation of Dynamical Gauge Fields,}
Phys. Rev. Lett. {\bf 111}, 110504 (2013).

\bibitem{Marcos14}
D. Marcos, P. Widmer, E. Rico, M. Hafezi, P. Rabl, U.-J. Wiese, and P. Zoller,
{\it Two-dimensional lattice gauge theories with superconducting quantum circuits,}
Ann. Phys. {\bf 351}, 634 (2014).

\bibitem{Hafezi13}
M. Hafezi, P. Adhikari, and J. M. Taylor, 
{\it Engineering three-body interaction and Pfaffian states in circuit QED systems,}
Phys. Rev. B {\bf 90}, 060503(R) (2014)

\bibitem{Brennen15}
G. K. Brennen, G. Pupillo, E. Rico, T. M. Stace, and D. Vodola,
{\it Loops and strings in a superconducting lattice gauge simulator},
Phys. Rev. Lett. {\bf 117}, 240504 (2016).

\bibitem{Lechner15}
W. Lechner, P. Hauke and P. Zoller,
{\it A quantum annealing architecture with all-to-all connectivity from local interactions,}
Sci. Adv. {\bf 1}, e1500838 (2015).

\bibitem{Chancellor16}
N. Chancellor, S. Zohren, and P. A. Warburton,
{\it Circuit design for multi-body interactions in superconducting quantum annealing system with applications to a scalable architecture,}
arXiv:1603:09521 (2016)

\bibitem{Leib16}
M. Leib, P. Zoller, and W. Lechner,
{\it A Transmon Quantum Annealer: Decomposing Many-Body Ising Constraints Into Pair Interactions,}
Quantum Sci. Technol. {\bf 1} 015008 (2016)

\bibitem{Chen14}
Y. Chen, P. Roushan, D. Sank, C. Neill, Erik Lucero, Matteo Mariantoni, R. Barends, B. Chiaro, J. Kelly, A. Megrant, J. Y. Mutus, P. J. J. O'Malley, A. Vainsencher, J. Wenner, T. C. White, Yi Yin, A. N. Cleland, and John M. Martinis,
{\it Simulating weak localization using superconducting quantum circuits}
Nat. Commun. {\bf 5}, 5184 (2014)

\bibitem{Barends15}
R. Barends, L. Lamata, J. Kelly, L. Garci'a-Alvarez, A. G. Fowler, A. Megrant, E. Jeffrey, T. C. White, D. Sank, J. Y. Mutus, B. Campbell, Yu Chen, Z. Chen, B. Chiaro, A. Dunsworth, I.-C. Hoi, C. Neill, P. J. J. O'Malley, C. Quintana, P. Roushan, A. Vainsencher, J. Wenner, E. Solano, and John M. Martinis,
{\it Digital quantum simulation of fermionic models with a superconducting circuit},
Nat. Commun. {\bf 6}, 7654 (2015).

\bibitem{Salathe15}
Y. Salath\`e, M. Mondal, M. Oppliger, J. Heinsoo, P. Kurpiers, A. Poto\v{c}nik, A. Mezzacapo, U. Las Heras, L. Lamata, E. Solano, S. Filipp, and A. Wallraff,
{\it Digital quantum simulation of spin models with circuit quantum electrodynamics},
Phys. Rev. X {\bf 5}, 021027 (2015).

\bibitem{Eichler15}
C. Eichler, J. Mlynek, J. Butscher, P. Kurpiers, K. Hammerer, T.?J. Osborne, and A. Wallraff,
{\it Exploring Interacting Quantum Many-Body Systems by Experimentally Creating Continuous Matrix Product States in Superconducting Circuits},
Phys. Rev. X {\bf 5}, 041044 (2015).

\bibitem{OMalley16}
P. J. J. O'Malley, R. Babbush, I. D. Kivlichan, J. Romero, J. R. McClean, R. Barends, J. Kelly, P. Roushan, A. Tranter, N. Ding, B. Campbell, Y. Chen, Z. Chen, B. Chiaro, A. Dunsworth, A. G. Fowler, E. Jeffrey, A. Megrant, J. Y. Mutus, C. Neill, C. Quintana, D. Sank, A. Vainsencher, J. Wenner, T. C. White, P. V. Coveney, P. J. Love, H. Neven, A. Aspuru-Guzik, and J. M. Martinis,
{\it Scalable Quantum Simulation of Molecular Energies}
Phys. Rev. X {\bf 6}, 031007 (2016).

\bibitem{Reed12}
M. D. Reed,	L. DiCarlo,	S. E. Nigg,	L. Sun,	L. Frunzio,	S. M. Girvin and R. J. Schoelkopf,
{\it Realization of three-qubit quantum error correction with superconducting circuits},
Nature {\bf 482}, 382 (2012).

\bibitem{Steffen13}
L. Steffen, Y. Salath\`e, M. Oppliger, P. Kurpiers, M. Baur, C. Lang, C. Eichler, G. Puebla-Hellmann, A. Fedorov, and A. Wallraff,
{\it Deterministic quantum teleportation with feed-forward in a solid state system},
Nature {\bf 500}, 319 (2013).

\bibitem{Barends14}
R. Barends,  J. Kelly,  A. Megrant,  A. Veitia,  D. Sank,  E. Jeffrey,  T. C. White,  J. Mutus,  A. G. Fowler,  B. Campbell,  Y. Chen,  Z. Chen,  B. Chiaro,  A. Dunsworth,  C. Neill,  P. O'Malley,  P. Roushan,  A. Vainsencher,  J. Wenner,  A. N. Korotkov,  A. N. Cleland, and J. M. Martinis,
{\it Superconducting quantum circuits at the surface code threshold for fault tolerance}
Nature {\bf 508}, 500 (2014)

\bibitem{Barends16}
R. Barends, A. Shabani, L. Lamata, J. Kelly, A. Mezzacapo, U. Las Heras, R. Babbush, A. G. Fowler, B. Campbell, Yu Chen, Z. Chen, B. Chiaro, A. Dunsworth, E. Jeffrey, E. Lucero, A. Megrant, J. Y. Mutus, M. Neeley, C. Neill, P. J. J. O'Malley, C. Quintana, P. Roushan, D. Sank, A. Vainsencher, J. Wenner, T. C. White, E. Solano, H. Neven, and John M. Martinis,
{\it Digitized adiabatic quantum computing with a superconducting circuit},
Nature {\bf 534}, 222 (2016)

\bibitem{Ofek16}
N. Ofek, A. Petrenko, R. Heeres, P. Reinhold, Z. Leghtas, B. Vlastakis, Y. Liu, L. Frunzio, S. Girvin, L. Jiang, M. Mirrahimi, M.H. Devoret, and R.J. Schoelkopf,
{\it Extending the lifetime of a quantum bit with error correction in superconducting circuits},
Nature, advance online publication (2016).

\bibitem{Fitzpatrick16}
M. Fitzpatrick, N. M. Sundaresan, A. C. Y. Li, J. Koch, and A. A. Houck,
{\it Observation of a dissipative phase transition in a one-dimensional circuit QED lattice},
arXiv:1607.06895 (2016).

\bibitem{Filipp09}
S. Filipp, P. Maurer, P. J. Leek, M. Baur, R. Bianchetti, J. M. Fink, M. G\"oppl, L. Steffen, J. M. Gambetta, A. Blais, and A. Wallraff,
{\it Two-Qubit State Tomography Using a Joint Dispersive Readout},
Phys. Rev. Lett. {\bf 102}, 200402 (2009).

\bibitem{DiCarlo10}
L. DiCarlo,	 M. D. Reed,	 L. Sun,	 B. R. Johnson,	 J. M. Chow,	 J. M. Gambetta,	 L. Frunzio,	 S. M. Girvin, M. H. Devoret, and  R. J. Schoelkopf,
{\it Preparation and measurement of three-qubit entanglement in a superconducting circuit}, 
Nature {\bf 467}, 574 (2010)

\bibitem{Roushan16}
P. Roushan, C. Neill, A. Megrant, Y. Chen, R. Babbush, R. Barends, B. Campbell, Z. Chen, B. Chiaro, A. Dunsworth, A. Fowler, E. Jeffrey, J. Kelly, E. Lucero, J. Mutus, P. J.J. O'Malley, M. Neeley, C. Quintana, D. Sank, A. Vainsencher, J. Wenner, T. White, E. Kapit, H. Neven, J. Martinis,
{\it Chiral groundstate currents of interacting photons in a synthetic magnetic field},
Nat. Phys. {\bf 13}, 146 (2017).

\bibitem{Abuwasib13}
M. Abuwasib, P. Krantz and P. Delsing,
{\it Fabrication of large dimension aluminum air-bridges for superconducting quantum circuits}
J. Vac. Sci. Technol. B {\bf 31}, 031601 (2013).

\bibitem{Chen14a}
Z. Chen, A. Megrant, J. Kelly, R. Barends, J. Bochmann, Yu Chen, B. Chiaro, A. Dunsworth, E. Jeffrey, J. Y. Mutus, P. J. J. O'Malley, C. Neill, P. Roushan, D. Sank, A. Vainsencher, J. Wenner, T. C. White, A. N. Cleland and John M. Martinis.
{\it Fabrication and characterization of aluminum airbridges for superconducting microwave circuits},
Appl. Phys. Lett. {\bf 104}, 052602 (2014).

\bibitem{Jin13}
J. Jin, D. Rossini, R. Fazio, M. Leib, and M. J. Hartmann,
{\it Photon solid phases in driven arrays of non-linearly coupled cavities},
Phys. Rev. Lett. {\bf 110}, 163605 (2013).
%
\bibitem{Neumeier13}  L. Neumeier, M. Leib, and M. J. Hartmann,
{\it Single Photon Transistor in Circuit Quantum Electrodynamics},
Phys. Rev. Lett. {\bf 111}, 063601 (2013).

\bibitem{Weimer10}
H. Weimer, M. M\"uller, I. Lesanovsky, P. Zoller, and H. P. B\"uchler,
{\it A Rydberg quantum simulator,}
Nat. Phys. {\bf 6}, 382 (2010).

\bibitem{Becker13}
D. Becker, T. Tanamoto, A. Hutter, F. L. Pedrocchi, and D. Loss,
{\it Dynamic generation of topologically protected self-correcting quantum memory}
Phys. Rev. A {\bf 87}, 042340 (2013).

\bibitem{Tanamoto13}
T. Tanamoto, V. M. Stojanovi\'{c}, C. Bruder, and D. Becker,
{\it Strategy for implementing stabilizer-based codes on solid-state qubits},
Phys. Rev. A {\bf 87}, 052305 (2013).

\bibitem{Mezzacapo14}
A. Mezzacapo, L. Lamata, S. Filipp, and E. Solano,
{\it Many-Body Interactions with Tunable-Coupling Transmon Qubits}
Phys. Rev. Lett. {\bf 113}, 050501 (2014).

\bibitem{Bombin07}
H. Bombin and M. A. Martin-Delgado,
{\it Optimal resources for topological two-dimensional stabilizer codes: Comparative study},
Phys. Rev. A 76, 012305 (2007).

\bibitem{Tomita14}
Y. Tomita and K. M. Svore,
{\it Low-distance surface codes under realistic quantum noise},
Phys. Rev. A {\bf 90}, 062320 (2014).

\bibitem{Gambetta15}
J. M. Gambetta, J. M. Chow, and M. Steffen,
{\it Building logical qubits in a superconducting quantum computing system},
arXiv:1510.04375 (2015).

\bibitem{Koch07}
J. Koch, T.M. Yu, J. Gambetta, A.A. Houck, D.I. Schuster, J. Majer, A. Blais, M.H. Devoret, S.M. Girvin, and R.J. Schoelkopf, 
{\it Charge-insensitive qubit design derived from the Cooper pair box},
Phys. Rev. A {\bf 76}, 042319 (2007).

\bibitem{Kafri16}
D. Kafri, Ch. Quintana, Y. Chen, A. Shabani, J. M. Martinis, and H. Neven
{\it Tunable inductive coupling of superconducting qubits in the strongly nonlinear regime},
arXiv:1606.08382 (2016).

\bibitem{Nakamura99}
Y. Nakamura, Yu. A. Pashkin, and J. S. Tsai,
{\it Coherent control of macroscopic quantum states in a single-Cooper-pair box},
Nature {\bf 398}, 786 (1999)

\bibitem{Bravyi08}
S. Bravyi, D. DiVincenzo, D. Loss, and B. Terhal,
{\it Simulation of Many-Body Hamiltonians using Perturbation Theory with Bounded-Strength Interactions,}
Phys. Rev. Lett. {\bf 101}, 070503 (2008). 

\bibitem{Koenig10}
R. K\"onig,
{\it Simplifying quantum double Hamiltonians using perturbative gadgets,}
Quant. Inf. Comp. {\bf 10}, 292 (2010).

\bibitem{Brell10}
C. G. Brell, S. Flammia, S.Bartlett, and A. Doherty,
{\it Toric codes and quantum doubles from two-body Hamiltonians,}
New J. Phys. {\bf 13}, 053039 (2011).

\bibitem{Schrieffer66}
J. R. Schrieffer and P. A. Wolff,
{\it Relation between the Anderson and Kondo Hamiltonians},
Phys. Rev. {\bf 149}, 491 (1966).

\bibitem{Niskanen07}
A. O. Niskanen, K. Harrabi, F. Yoshihara, Y. Nakamura, S. Lloyd, and J. S. Tsai,
{\it Quantum Coherent Tunable Coupling of Superconducting Qubits,}
Science {\bf 316}, 723 (2007).

\bibitem{Masluk12}
N. A. Masluk, I. M. Pop, A. Kamal, Z. K. Minev, and M. H. Devoret
{\it Microwave Characterization of Josephson Junction Arrays: Implementing a Low Loss Superinductance,}
Phys. Rev. Lett. {\bf 109}, 137002 (2012). 

\bibitem{Bell12}
M. T. Bell, I. A. Sadovskyy, L. B. Ioffe, A. Yu. Kitaev, and M. E. Gershenson,
{\it Quantum Superinductor with Tunable Nonlinearity,}
Phys. Rev. Lett. {\bf 109}, 137003 (2012).

\bibitem{Samkharadze16}
N. Samkharadze, A. Bruno, P. Scarlino, G. Zheng, D. P. DiVincenzo, L. DiCarlo, and L. M. K. Vandersypen
{\it High-Kinetic-Inductance Superconducting Nanowire Resonators for Circuit QED in a Magnetic Field},
Phys. Rev. Applied {\bf 5}, 044004 (2016).

\bibitem{Bultink16}
C. C. Bultink, M. A. Rol, T. E. O'Brien, X. Fu, B. C. S. Dikken, C. Dickel, R. F. L. Vermeulen, J. C. de Sterke, A. Bruno, R. N. Schouten, and L. DiCarlo,
{\it Active resonator reset in the nonlinear dispersive regime of circuit QED},
Phys. Rev. Applied {\bf 6}, 034008 (2016).
	
\bibitem{Chen16}
Z. Chen, J. Kelly, C. Quintana, R. Barends, B. Campbell, Yu Chen, B. Chiaro, A. Dunsworth, A. G. Fowler, E. Lucero, E. Jeffrey, A. Megrant, J. Mutus, M. Neeley, C. Neill, P. J. J. O'Malley, P. Roushan, D. Sank, A. Vainsencher, J. Wenner, T. C. White, A. N. Korotkov, and J. M. Martinis,
{\it Measuring and Suppressing Quantum State Leakage in a Superconducting Qubit},
Phys. Rev. Lett. {\bf 116}, 020501 (2016).

\bibitem{Quintana14}
C. M. Quintana, A. Megrant, Z. Chen, A. Dunsworth, B. Chiaro, R. Barends, B. Campbell, Yu Chen, I.-C. Hoi, E. Jeffrey, J. Kelly, J. Y. Mutus, P. J. J. O'Malley, C. Neill, P. Roushan, D. Sank, A. Vainsencher, J. Wenner, T. C. White, A. N. Cleland, and J. M. Martinis,
{\it Characterization and reduction of microfabrication-induced decoherence in superconducting quantum circuits},
Appl. Phys. Lett. {\bf 105}, 062601 (2014)

\bibitem{Chow15}
J. M. Chow, S. J. Srinivasan, E. Magesan, A. D. C\'{o}rcoles, D. W. Abraham, J. M. Gambetta, and M. Steffen,
{\it Characterizing a four-qubit planar lattice for arbitrary error detection},
Proc. SPIE  9500, Quantum Information and Computation XIII, 95001G (2015)

\bibitem{Trebst07}
S. Trebst, P. Werner, M. Troyer, K. Shtengel, and C. Nayak,
{\it Breakdown of a Topological Phase: Quantum Phase Transition in a Loop Gas Model with Tension},
Phys. Rev. Lett. {\bf 98}, 070602 (2007).

\bibitem{Dusuel11}
S. Dusuel, M. Kamfor, R. Or\'{u}s, K. P. Schmidt, and J. Vidal,
{\it Robustness of a Perturbed Topological Phase},
Phys. Rev. Lett. {\bf 106}, 107203 (2011).

\bibitem{You10}
J. Q. You, X.-F. Shi, X. Hu, and F. Nori,
{\it Quantum emulation of a spin system with topologically protected ground states using superconducting circuits},
Phys. Rev. B {\bf 81}, 014505 (2010).

\bibitem{Brecht16}
T. Brecht, W. Pfaff, Ch. Wang, Y. Chu, L. Frunzio, M. H. Devoret and R. J. Schoelkopf,
{\it Multilayer microwave integrated quantum circuits for scalable quantum computing}
npj Quantum Information {\bf 2}, 16002 (2016).

\bibitem{Bejamin16}
J.H. B\'{e}janin, T.G. McConkey, J.R. Rinehart, C.T. Earnest, C.R.H. McRae, D. Shiri, J.D. Bateman, Y. Rohanizadegan, B. Penava, P. Breul, S. Royak, M. Zapatka, A.G. Fowler, and M. Mariantoni,
{\it The Quantum Socket: Three-Dimensional Wiring for Extensible Quantum Computing},
Phys. Rev. Applied {\bf 6}, 044010 (2016).

\bibitem{Schmitt14}
V. Schmitt, X. Zhou, K. Juliusson, B. Royer, A. Blais, P. Bertet, D. Vion, and D. Esteve,
{\it Multiplexed readout of transmon qubits with Josephson bifurcation amplifiers},
Phys. Rev. A {\bf 90}, 062333 (2014)

\bibitem{Chen12}
Yu Chen, D. Sank, P. O'Malley, T. White, R. Barends, B. Chiaro, J. Kelly, E. Lucero, M. Mariantoni, A. Megrant, C. Neill, A. Vainsencher, J. Wenner, Y. Yin, A. N. Cleland and John M. Martinis
{\it Multiplexed dispersive readout of superconducting phase qubits}
Appl. Phys. Lett. {\bf 101}, 182601 (2012)

\bibitem{Jerger12}
M. Jerger, S. Poletto, P. Macha, U. H\"ubner, E. Il'ichev and A. V. Ustinov,
{\it Frequency division multiplexing readout and simultaneous manipulation of an array of flux qubits},
Appl. Phys. Lett. {\bf 101}, 042604 (2012)

\bibitem{Reed10}
M. D. Reed, B. R. Johnson, A. A. Houck, L. DiCarlo, J. M. Chow, D. I. Schuster, L. Frunzio, and R. J. Schoelkopf
{\it Fast reset and suppressing spontaneous emission of a superconducting qubit},
Appl. Phys. Lett. {\bf 96}, 203110 (2010).

\bibitem{Jeffrey14}
Evan Jeffrey, Daniel Sank, J. Y. Mutus, T. C. White, J. Kelly, R. Barends, Y. Chen, Z. Chen, B. Chiaro, A. Dunsworth, A. Megrant, P. J. J. O'Malley, C. Neill, P. Roushan, A. Vainsencher, J. Wenner, A. N. Cleland, and John M. Martinis,
{\it Fast Accurate State Measurement with Superconducting Qubits},
Phys. Rev. Lett. {\bf 112}, 190504 (2014).

\bibitem{Pechal16}
M. Pechal, J.-C. Besse, M. Mondal, M. Oppliger, S. Gasparinetti, A. Wallraff,
{\it Superconducting switch for fast on-chip routing of quantum microwave fields}
Phys. Rev. Applied {\bf 6}, 024009 (2016).

\bibitem{Asaad15}
S. Asaad, C. Dickel, S. Poletto, A. Bruno, N. K. Langford, M. A. Rol, D. Deurloo, L. DiCarlo,
{\it Independent, extensible control of same-frequency superconducting qubits by selective broadcasting},
Npj Quantum Information {\bf 2}, 16029 (2016).

\bibitem{Hamma08}
A. Hamma and D. Lidar,
{\it Adiabatic preparation of topological order},
Phys. Rev. Lett. {\bf 100}, 030502 (2008).

\bibitem{Kapit14}
E. Kapit, M. Hafezi, and S. H. Simon,
{\it Induced Self-Stabilization in Fractional Quantum Hall States of Light},
Phys. Rev. X {\bf 4}, 031039 (2014).

\bibitem{Bardyn16}
C.-E. Bardyn and T. Karzig,
{\it Exponential lifetime improvement in topological quantum memories}
Phys. Rev. B {\bf 94}, 094303 (2016).


\bibitem{Hofheinz09}
M Hofheinz, H. Wang, M. Ansmann, Radoslaw C. Bialczak, Erik Lucero, M. Neeley, A. D. O'Connell, D. Sank, J. Wenner, J. M. Martinis, and A. N. Cleland,
{\it Synthesizing arbitrary quantum states in a superconducting resonator},
Nature {\bf 459}, 546 (2009).

\bibitem{Han07}
Y.-J. Han, R. Raussendorf, and L.-M. Duan,
{\it Scheme for Demonstration of Fractional Statistics of Anyons in an Exactly Solvable Model}
Phys. Rev. Lett. {\bf 98}, 150404 (2007).

\bibitem{Bombin06}
H. Bombin and M. A. Martin-Delgado,
{\it Topological Quantum Distillation},
Phys. Rev. Lett. {\bf 97}, 180501 (2006).

\bibitem{Goldman14}
N. Goldman and J. Dalibard,
{\it Periodically Driven Quantum Systems: Effective Hamiltoninas and Engineered Gauge Fields,}
Phys. Rev. X {\bf 4}, 031027 (2014).

\bibitem{solid_NMR}
I. Scholz, B. H. Meier, and M. Ernst,
{\it Operator-based triple-mode Floquet theory in solid-state NMR}, 
J. Chem. Phys. {\bf 127}, 204504 (2007).

\bibitem{James00}
D. F. V. James,
{\it Quantum Computation with hot and cold ions: An assessment of proposed schemes},
Fortschritte der Physik {\bf 48}, 823 (2000).



\end{thebibliography}
\end{document}